\documentclass[aps,pre,longbibliography]{revtex4-2}
\usepackage{graphicx}
\usepackage{fancybox}
\usepackage{longtable}
\usepackage{amsmath,amssymb,bm,amsthm}
\usepackage[T1]{fontenc}
\usepackage[utf8]{inputenc}
\usepackage{xcolor}
\def\beq{\begin{equation}}
\def\eeq{\end{equation}}

\def\<{\langle}
\def\>{\rangle}

\newcommand*{\addheight}[2][.5ex]{%
  \raisebox{0pt}[\dimexpr\height+(#1)\relax]{#2}%
}

\begin{document}

\title{
Multicritical Bifurcation and First-order Phase Transitions in a Three-dimensional 
Blume-Capel Antiferromagnet 
}

\date{\today}

\author{Daniel Silva$^1$, Gloria M.\ Buend{\'\i}a$^2$, and Per Arne Rikvold$^{1,3}$}
\email[Corresponding author: ]{p.a.rikvold@fys.uio.no}
\affiliation{$^1$ Department of Physics, Florida State University, 
Tallahassee, FL 32306-4350, USA \\
$^2$Department of Physics, Universidad Sim{\'o}n Bol{\'{\i}}var, 
Caracas 1080, Venezuela\\
$^3$PoreLab, NJORD Centre, Department of Physics, University of Oslo, 
P.O.\ Box 1048 Blindern, 0316 Oslo, Norway}

\begin{abstract}
We present a detailed study by Monte Carlo simulations and finite-size scaling analysis 
of the phase diagram and ordered bulk phases for the three-dimensional 
Blume-Capel antiferromagnet in the space of temperature and magnetic and crystal fields 
(or two chemical potentials in an equivalent lattice-gas model with two particle species 
and vacancies). 
The phase diagram consists of surfaces of second- and first-order transitions that 
enclose a ``volume'' of ordered phases in the phase space. 
At relatively high temperatures, these surfaces join smoothly along a line of tricritical points, 
and at zero magnetic field we obtain good agreement with known values for 
tricritical exponent ratios [Y.\ Deng and H.W.J.~Bl{\"o}te, Phys.\ Rev. E {\bf 70}, 0456111 
(2004)]. 
In limited field regions at lower 
temperatures (symmetric under reversal of the magnetic field), 
the tricritical line for this three-dimensional model 
bifurcates into lines of critical endpoints and  critical points, 
connected by a surface of weak first-order transitions {\it inside} the region of ordered phases. 
This phenomenon is not seen in the two-dimensional version of the same model. 
We confirm the location of the bifurcation as previously reported 
[Y.L.\ Wang and J.D.\ Kimel, J.\ Appl.\ Phys.\ {\bf 69}, 6176 (1991)], 
and we identify the phases separated by this first-order surface as 
antiferromagnetically (three-dimensional checker-board) 
ordered with different vacancy densities. We visualize the phases by 
real-space snapshots and by structure factors in the three-dimensional space of wave vectors. 
\end{abstract}

\maketitle

\section{Introduction}
\label{sec:I}

The spin-1 (three-state) Ising model with bilinear and single-ion anisotropy is known as the 
Blume-Capel (BC) model. 
It was introduced independently in 1966 by Blume \cite{BLUME66} and Capel \cite{CAPE66} 
to describe certain magnetic phase transitions. 
Over the ensuing six decades, the complex phase behavior of the BC model 
(often in an equivalent lattice-gas formulation with two particle species and vacancies, 
or in generalizations to various lattices 
and with additional interactions \cite{BLUME71,COLL89,JZHA95B})
has been used to explore critical and multicritical 
properties in a variety of physical systems. Among these applications are  
the $\lambda$ transition and phase separation in He$^3$-He$^4$ mixtures \cite{BLUME71}, 
dielectric properties of nanowires \cite{BENH2018}, 
phase behavior of superionic liquids in nanoporous media 
\cite{DUDKA2016, DUDKA2019,DUDKA2021}, 
two-component adsorption from liquids or gases \cite{COLL89,JZHA95B,SILVA2019,FEFE19}, 
ternary steel alloys \cite{LARA22},
and exotic ``nuclear pasta'' phases thought to exist in neutron stars \cite{HASN13}, 
to mention just a few.

The rich phase diagram of the BC model, 
with surfaces of first- and second-order transitions that are smoothly joined along lines of 
tricritical points \cite{LAW84}, or along lines of critical endpoints where the second-order 
surface meets the first-order surface at a finite angle \cite{COLL88,WILD97}, 
has also served as a test bed for many theoretical and numerical techniques. 
These include mean-field approaches, finite-size scaling, and various Monte Carlo 
techniques in two and three dimensions. 
Several of these studies are relevant to the present work and are cited below.   

In three dimensions (3D), 
the antiferromagnetic (AFM) BC model on a simple cubic lattice presents an 
intriguing feature:  
the line of tricritical points decomposes at low temperatures into a line of critical endpoints
and one of critical points. These lines of critical endpoints and critical points are 
connected by an extension of the first-order transition surface into the ordered-phase region of the phase diagram. 
This phenomenon was first observed by mean-field theory \cite{WANG1976} and 
Monte Carlo simulations \cite{KIME91A}. 
However, finite-size scaling analyses of data from large-scale 
numerical transfer-matrix calculations and 
Monte Carlo simulations conclusively show that it does not occur in the 2D, 
square-lattice version of the same model. 
Instead, the tricritical line in the 2D model continues unbroken all the way down to zero 
temperature \cite{KIME91}. 
This qualitative difference between the behaviors in two and three dimensions 
has been attributed to the presence of large fluctuations in the 2D case \cite{KIME91}. 

Despite the enduring popularity of applications of the BC model to various physical 
and chemical systems, ranging from nuclear astrophysics to metallurgy, 
we are not aware of any further study of multicriticality in the 3D, AFM BC model.  
Therefore, we here present a 
comprehensive Monte Carlo study on simple cubic lattices up to 
$32^3$ sites of the phase diagram in the space of 
magnetic field, crystal field, and temperature, 
both on a global scale and on a fine scale in the phase region of the decomposition.  
This enables us to detect and describe three distinct phases of different order and density 
in the region of phase space beyond the decomposition point.

The remainder of this paper is organized as follows. 
In Sec.~\ref{sec:mod} we define the model, describe the Monte Carlo method, and 
define the order parameters and other quantities that are extracted from the simulated 
time series and used in our finite-size scaling analysis. 
In Sec.~\ref{sec:large} we discuss the phase diagram on a large scale, 
including the ground-state diagram (\ref{sec:grd}) and the 
finite-temperature phase diagram (\ref{sec:finite}) that consists of second-order (\ref{sec:secondX}) 
and first-order (\ref{sec:first}) transitions and 
the line of tricritical points (\ref{sec:3Cline}). 
In Sec.~\ref{sec:Decomp} we study the bifurcation region in detail, including second-order 
(\ref{sec:second}) and first-order (\ref{sec:firstF}) transitions and the 
decomposed region (\ref{sec:decompX}). 
The structures of the distinct phases identified in the phase region beyond the bifurcation point 
are investigated in Sec.~\ref{sec:Phas} with snapshots (\ref{sec:Snap}) and 
static structure factors (\ref{sec:SQ}). 
A summary and our conclusions are given in Sec.~\ref{sec:Conc}.

\section{Model, Simulation, and Analysis Methods}
\label{sec:mod}

The Blume-Capel model is defined by the Hamiltonian,
\begin{equation}\label{eq:BLCP}
\mathcal {H}_{\text{BC}} 
    = -J \sum_{\langle i,j \rangle} s_i s_j + D \sum_i s_{i}^2  - H \sum_{i} s_i \;,
\end{equation}
where the ``spin'' variables $s_i \in \{-1,0,1\}$, $\sum_{\langle i,j \rangle}$ runs over all nearest-neighbor (nn) pairs, and  $\sum_i$ runs over all lattice sites. 
$J$ is the exchange parameter, here chosen 
negative to favor AFM (checker-board) ordering on two interpenetrating sublattices. 
$D$ is the ``crystal'' 
field that distinguishes between $s_i = 0$ and $\pm 1$, and $H$ is an external ``magnetic'' field. 
For simplicity, we 
define the dimensionless parameters, $d=D/|J|$, $h=H/|J|$, and $t=T/|J|$ and take Boltzmann's 
constant as unity. The general three-state Ising model can be equivalently formulated as a 
lattice-gas model with the two nonzero values of $s_i$ representing two different particle types, 
A and B, and $s_i = 0$ representing vacancies. The transformation equations can be 
found in, e.g., \cite{COLL88,SILVA2019}. They yield the lattice-gas interaction energies, 
$\phi_{\rm XY}$, for the BC model as 
$\phi_{\rm AA} = \phi_{\rm BB} = J$ and $ \phi_{\rm AB} = -J$. 

We perform equilibrium Monte Carlo simulations in 3D on a simple cubic lattice of 
size $V = L \times L \times L$ with $L$ between 12 and 32, with periodic boundary conditions. 
To facilitate equilibration, direct transitions are allowed between all three spin states at 
randomly chosen lattice sites. 
The acceptance probability of a proposed transition, $s_i \rightarrow {s_i}'$, is given by the 
corresponding energy change, $\Delta E = E' - E$, by the Metropolis 
algorithm \cite{METR53},
\beq
P(s_i \rightarrow {s_i}') = {\rm Min} [1, e^{- \Delta E  / T}] \;.
\label{eq:met}
\eeq  

The order parameters of interest are the staggered magnetization $m_{s}$, 
the magnetization $m$, and the density $\rho$, all per unit volume: 

\begin{equation}
m_{s}=\frac{1}{V} \sum_{i} (-1)^{i}s_i 
\label{eq:ms}
\end{equation}
with $i$ even on one sublattice and odd on the other,
\begin{equation}\label{eq:DEFO}
m= \frac{1}{V} \sum_{i} s_i \;,
\end{equation}
\begin{equation}
\rho=\frac{1}{V} \sum_{i} s_{i}^2 \;.
\label{eq:rho}
\end{equation}

Monte Carlo time series of up to $9 \times 10^6$ Monte Carlo Steps per Site (MCSS) 
for the largest systems, with the first third used for equilibration and averages calculated
 over evenly spaced samples from the remaining two thirds.
The results were analyzed with standard methods \cite{JANK08}, 
including finite-size scaling \cite{PRIV90B} 
for the susceptibilities associated with each order parameter $\mathcal O$  \cite{JANK08}, 
\begin{equation}
\label{eq:susc2}
\chi(\mathcal{O})=  \cfrac{V}{T} \langle ( \mathcal{O} - \langle \mathcal{O}\rangle)^2 \rangle \;,
\end{equation}
and fourth-order Binder cumulants \cite{BIND81}, 
\beq
U_4(\mathcal{O}) = 1 - \frac{\langle( \mathcal{O} - \langle \mathcal{O} \rangle )^4  \rangle}
{3 \langle( \mathcal{O} - \langle \mathcal{O} \rangle )^2  \rangle^2} \;.
\label{eq:Binder4}
\eeq
In both cases, the angled brackets indicate averages over samples. 

The Binder cumulant
measures the non-gaussianity of the probability distribution of an observable, and it is  a 
valuable tool to find critical points and to characterize the nature of a phase transition. 
In the proximity of a second-order phase transition, as the system goes from the ordered to
 the disordered phase, the two peaks of the probability distribution merge. 
At the critical point, $U_4$ approaches a fixed value, $U_4^{*} \leq 2/3$, as a function of 
$L$. Therefore, the critical point can be identified as the crossing point of the 
Binder cumulants for different system sizes. 
The value of $U_4^{*}$ is weakly universal 
in the sense that, for a transition in a given universality class, it may depend on the  
boundary conditions and any anisotropy of the interactions \cite{SELK05}. 
Reference values quoted in the present paper are those appropriate for 
the 3D Ising class on the simple cubic lattice with 
periodic boundary conditions and isotropic interactions.

For more accurate location of first-order phase transitions, we also 
consider probability distributions $P(\mathcal{O})$ and 
their associated free-energy densities \cite{JLEE90, JLEE91}. 
Ordered and disordered phases at low, finite temperatures are investigated with 
snapshot images and 3D static structure factors. Further details of the various methods 
are given below as needed.

\section{Large-scale features of the phase diagram}
\label{sec:large}

\subsection{Ground-state diagram}
\label{sec:grd}

\begin{figure}
  \begin{center}
  \hspace*{-0mm}
\includegraphics[scale=0.6,width=10cm,height=8.8cm]{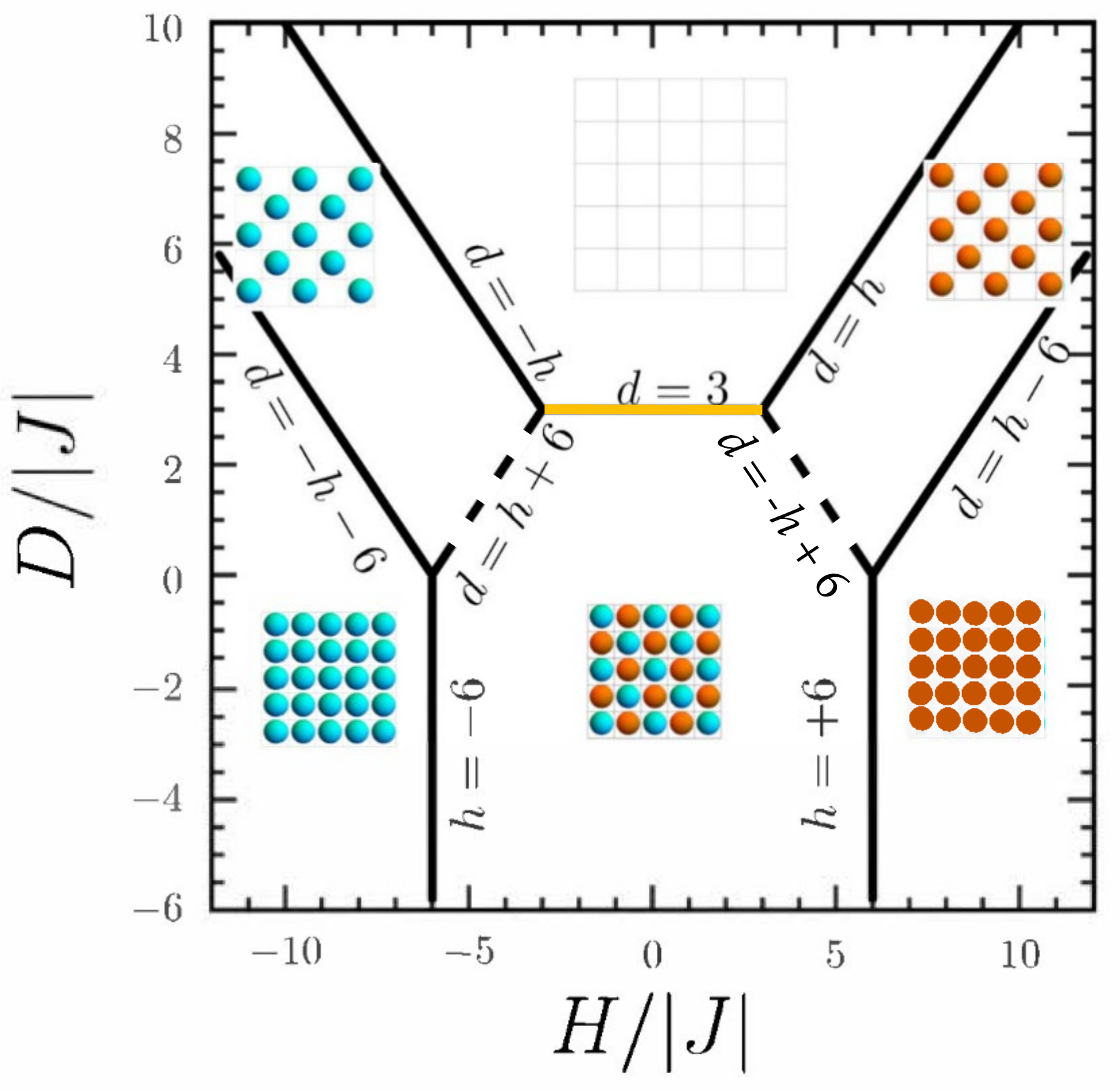}
  \end{center}
  \vspace*{-10mm}
  \caption{The ground-state diagram for the 3D, AFM Blume-Capel model in the $(h, d)$ 
parameter plane. The insets show 2D slices in the crystallographic 
(100) plane of the 3D ground states. 
Red (blue) points indicate +1 ($-1$) spins, and $s_i = 0$ is represented by empty sites.
At nonzero temperatures, the yellow, solid line continues as a surface of 
first-order phase transitions, while the black, solid lines continue as tunnel-shaped 
surfaces of second-order transitions, as shown in Fig.~\ref{fig:coasegbc}. 
The dashed lines continue as noncritical crossovers between the corresponding pairs of 
ordered phases \cite{COLL88,PAWL09}. 
}
  \label{fig:afmgs_bc}
\end{figure}

We first construct the ground-state diagram, 
which  can be thought of as the foundation of the finite-temperature phase diagram. 
It is obtained by calculating all the configurations of the unit cell for the different sets of parameters 
and selecting the ones with the lowest energy in each case. The equations for the boundaries 
between phase regions are found by pairwise equating the ground-state energies 
of the adjoining phases. This diagram is shown in 
Fig.~\ref{fig:afmgs_bc}. Except for numerical constants in the equations for the 
phase boundaries, it is identical to the one for the 2D 
version of the model \cite{KIME91,SILVA2019}.

\subsection{Finite-temperature phase diagram}
\label{sec:finite}

\begin{figure}
  \begin{center}
\includegraphics[scale=0.6,width=16cm,height=14cm]{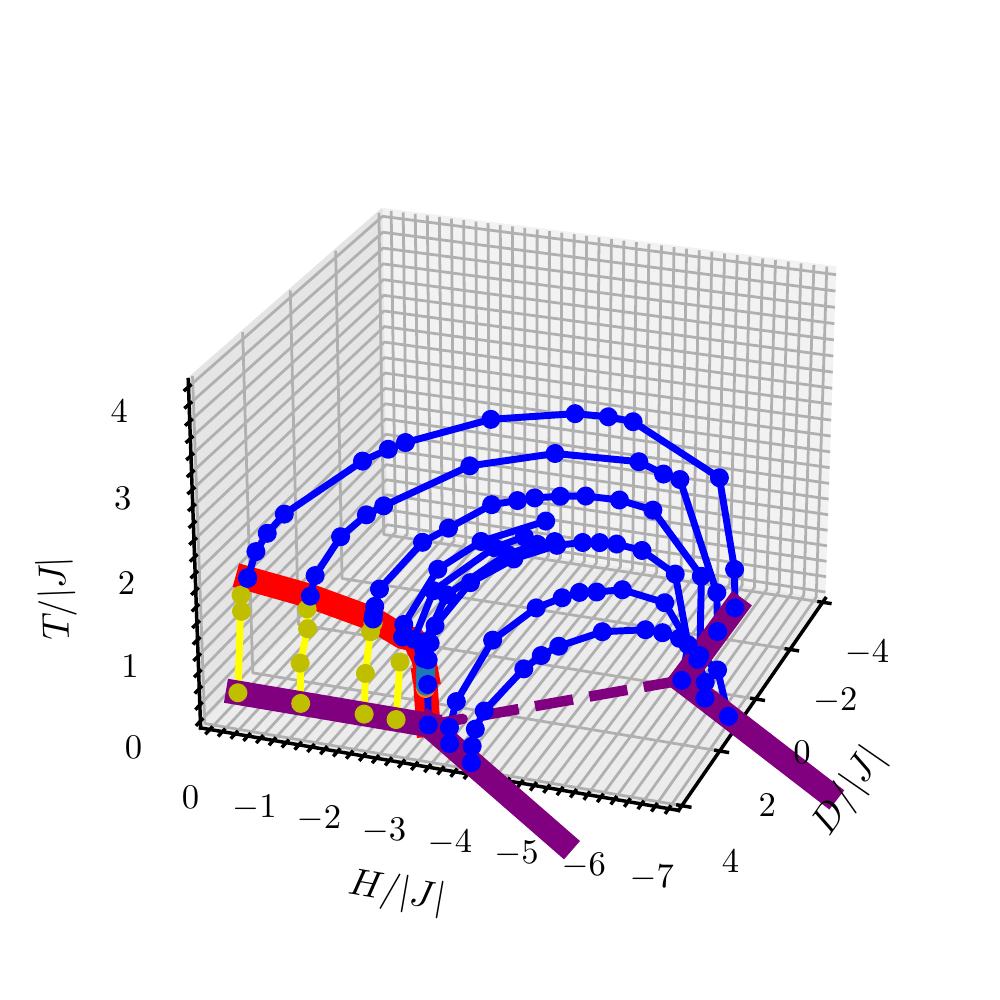}
  \end{center}
  \vspace*{-12mm}
  \caption{A large-scale view of the finite-temperature phase diagram, shown in the $\{ h,d,t \}$ space. The ``tunnel'' formed by blue points and lines represents a surface of second-order 
phase transitions, which smoothly joins a surface of first-order transitions (yellow) 
along a line of tricritical points (red). 
Phase boundaries from the ground-state diagram in Fig.~\ref{fig:afmgs_bc} ($T = 0$) 
are shown in purple. 
In order to clearly display the first-order surface and tricritical line, this 
image is viewed in the positive $h$ direction and the negative $d$ and $t$ directions 
from a phase point near $(h,d,t) = (-8, +6, +3)$. 
At this scale, the decomposition of the tricritical line in the limited parameter range 
$|h| \alt 3, d \alt 3, t < 0.57$ is not visible. 
It is discussed and shown in detail in Sec.~\ref{sec:decompX}. 
}
  \label{fig:coasegbc}
\end{figure}

To facilitate the further reading of this paper, we show in Fig.~\ref{fig:coasegbc} 
a large-scale view of the finite-temperature phase diagram. 
The yellow points represent a surface of first-order phase transitions,
which smoothly joins a surface of second-order transitions 
represented by blue points. The line, along which these surfaces join smoothly, 
consists of tricritical points \cite{LAW84}, indicated in red. 
How these surfaces and line were determined from our Monte Carlo data 
is outlined below. 

At this large scale, the phase diagram 
appears topologically identical to that for the 2D 
version of the model \cite{KIME91}. 
The decomposition of the tricritical line, mentioned in Sec.~\ref{sec:I} and discussed 
in detail in Secs.~\ref{sec:Decomp} and \ref{sec:Phas}, is confined to the phase region of 
$|h| \in [2.94,3] $, $d \in [+2.98,+3]$, and $t < 0.57$,
which is is too small to be visible on the scale of this figure.

\subsubsection{Second-order transitions}
\label{sec:secondX}

The points on the large, tunnel-like surfaces of 
second-order phase transitions in Fig.~\ref{fig:coasegbc} 
were located in standard fashion \cite{JANK08}.  
One of the three fields was scanned through the expected transition while the other two 
were kept constant. The transition point 
along the scan line was located by 
maxima of the susceptibility $\chi(m_s)$ and/or $\chi(\rho)$ and 
crossings of Binder cumulants for different $L$ in the range $12,...,32$. 
Even without $L$-extrapolations, this method 
yielded error bars smaller than the symbol size in Fig.~\ref{fig:coasegbc}. 
The order of the transition was ascertained by checking the power-law divergence of 
$\chi  \sim L^{\gamma/\nu}$ for compatibility of the observed $\gamma/\nu$ with the 
expected value of approximately $1.964$ for the 3D Ising universality class \cite{HASE10,RON17}. 
The weakly universal \cite{SELK05} Binder-cumulant crossing values were also checked to be 
in the vicinity of the expected value of 0.466 \cite{HASE10,FERR18,XU20}. 

\subsubsection{First-order transitions}
\label{sec:first}

The surface of first-order transitions is confined to a narrow range in $d$ between 
2.84 and 3 \cite{DESE97,DENG04,ZIER15}. We therefore located the points on this surface by 
scanning $d$ in this range at constant $h$ and $t$. 
As seen from their definition in Eq.~(\ref{eq:susc2}), the susceptibilities are proportional to the variance of the corresponding order-parameter distribution. When this distribution becomes 
sharply bimodal with $L$-independent peak separation as $L \rightarrow \infty$ at a first-order transition, the only remaining 
$L$-dependence in the susceptibilities is contained in the prefactor, $V = L^3$. 
We found the observation of this divergence in $\chi(\rho)$ sufficient to locate these points 
with an accuracy smaller than the symbol size.

\subsubsection{Tricritical line}
\label{sec:3Cline}

\begin{figure}[t]
    \begin{center}
    \hspace*{-10mm}
    \includegraphics[width=.35\paperwidth]{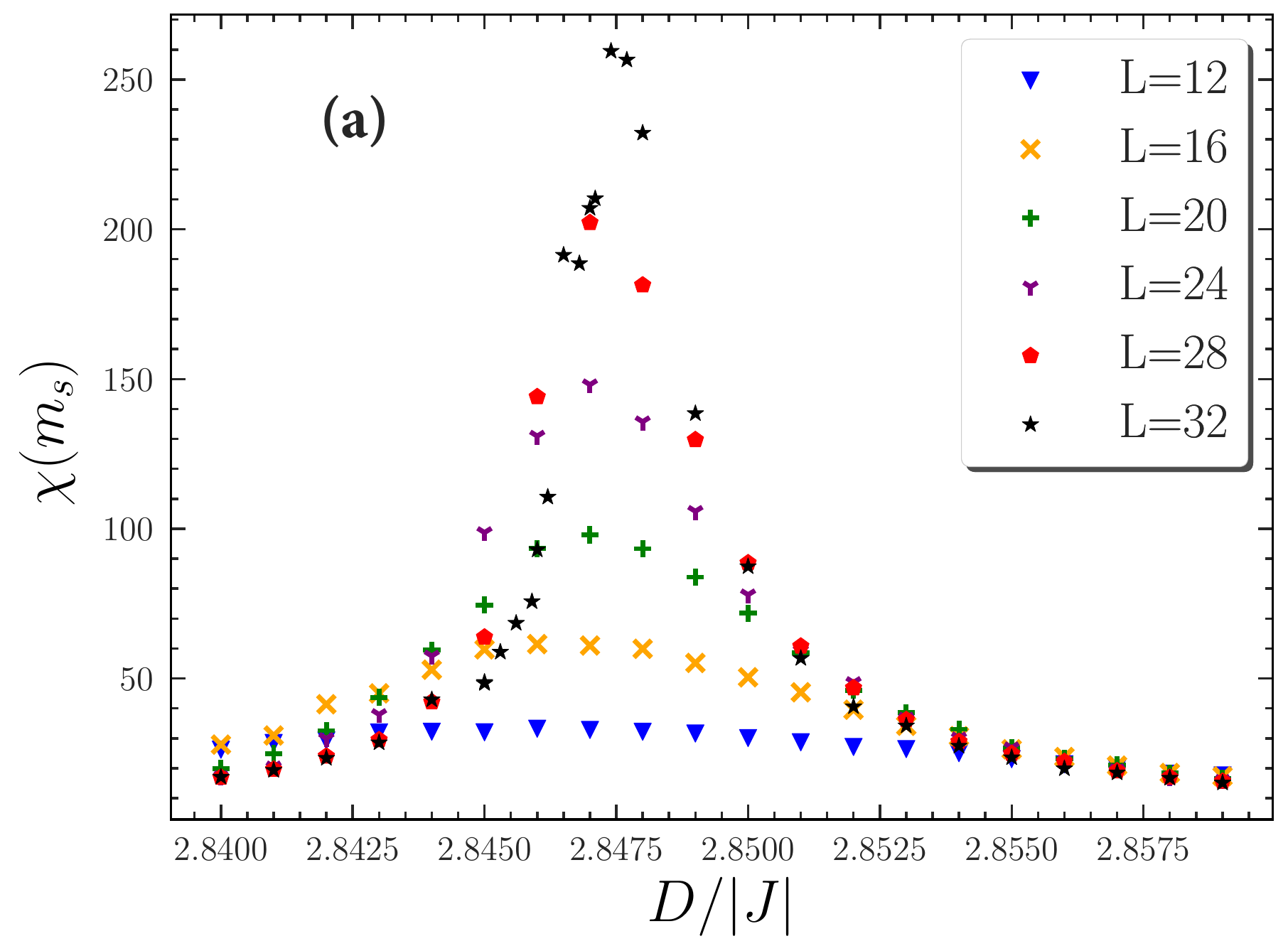}
    \includegraphics[width=.35\paperwidth]{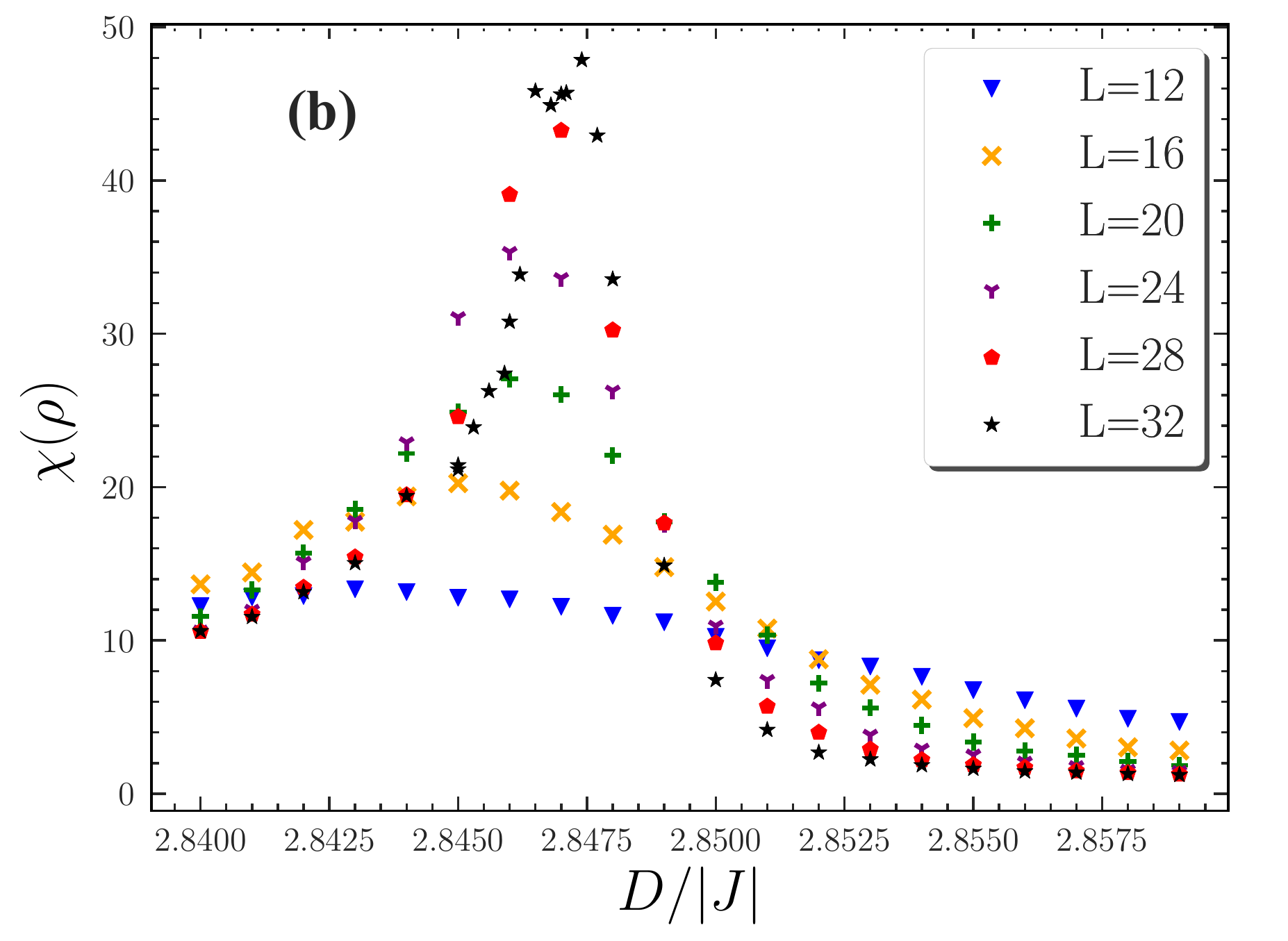}
    \includegraphics[width=.4\paperwidth]{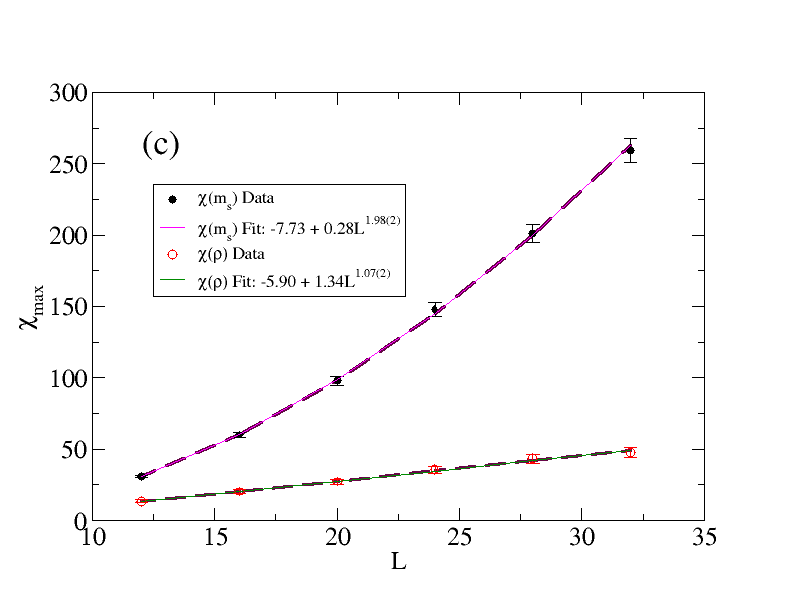}
  \end{center}
    \caption{Dependence on $d$ at constant $h=0$ and $t=1.402$ 
    of the susceptibilities  for $L = 12$, ..., 32. (a): $\chi(m_s)$ and (b): $\chi(\rho)$. 
     (c): Scaling plots of the susceptibility maxima vs $L$. The estimated values 
    of the scaling exponents, $\gamma/\nu$, are $1.98(2)$ and $1.07(2)$, respectively. 
    These are close to the expected tricritical values of 2 for $m_s$ and 1 for $\rho$ 
     \cite{DENG04}. 
    Dashed lines in the background are guides to the eye of form $a+bL^2$ and $a+bL$, 
    respectively.
    See discussion in the text. 
    }
    \label{fig:3CPC}
  \end{figure}

\begin{figure}[h]
    \begin{center}
    \hspace*{-10mm}
    \includegraphics[width=.35\paperwidth]{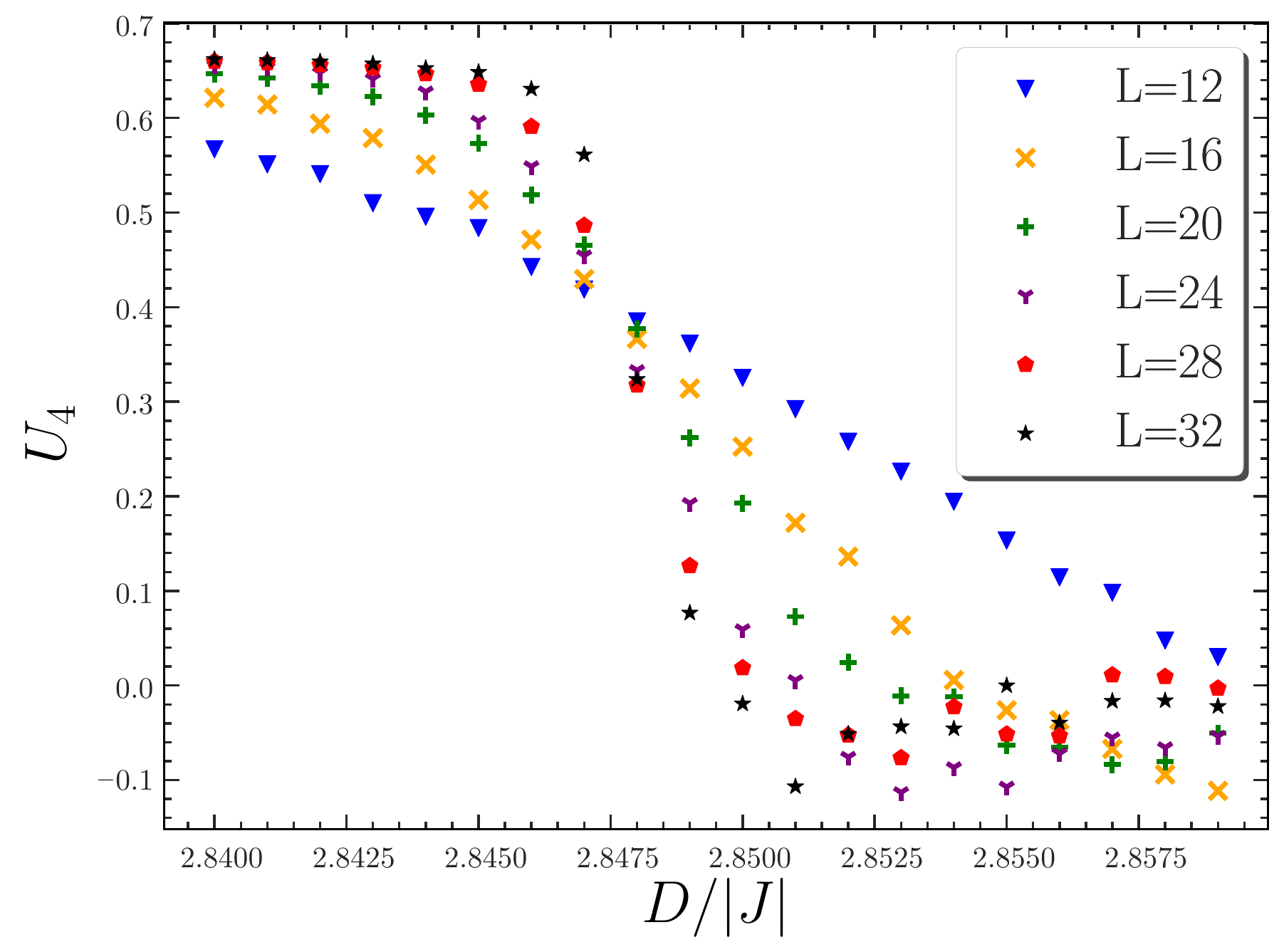}
  \end{center}
    \caption{Dependence on $d$ at constant $h=0$ and $t=1.402$ 
    of the fourth-order Binder cumulant $U_4(m_s)$  for $L = 12$, ..., 32. 
   The crossing value of the three largest systems, $U_4^* = 0.325(5)$, 
is  in excellent agreement with the expected tricritical value of $1/3$ \cite{DENG04}. 
    }
    \label{fig:3CPC_Binder}
  \end{figure}

At $h=0$, the tricritical points for the ferromagnetic and the antiferromagnetic BC models 
coincide by symmetry. 
For the ferromagnet, this point has been estimated by Monte Carlo simulations to be 
at $(d_t,t_t) = (2.84479(30),1.4182(55))$ \cite{DESE97,ZIER15} or $(2.848(1),1.4019(2))$ 
\cite{DENG04}, respectively. To verify our ability to reliably locate tricritical points 
and to ``anchor'' the tricritical line for $h \ne 0$, 
we performed scans in $d$ at constant $h=0$ and $t=1.402$. In Fig.~\ref{fig:3CPC} we show 
the susceptibilities, $\chi(m_s)$ and $\chi(\rho)$, and scaling plots 
of their maxima vs $L$. The 
slopes of the fitted lines are 1.98(2) and 1.07(2), respectively. These values are 
close to the theoretically expected values of 2 and 1, corresponding to 
$\gamma/\nu$ for perturbations nonparallel and parallel to the critical surface, 
respectively \cite{DENG04}. 
To account for relatively large, $L$-independent background terms, the fits were performed as weighted, nonlinear 3-parameter fits with error bars estimated as 
proportional to the variables. 
Extrapolations of $d(L)$ with respect to $1/L$ yield $d_t = 2.849(1)$, in good agreement with 
\cite{DENG04}.  The fourth-order Binder cumulant for $m_s$ is shown vs $d$ in 
Fig.~\ref{fig:3CPC_Binder}. The observed crossing values,  $d_t = 2.84760(5)$ and 
$U_4^* = 0.325(5)$ are close to the expected tricritical values from \cite{DENG04}.

\begin{figure}[htp]
    \begin{flushleft}
    \hspace*{-10mm}
    \includegraphics[width=.4\paperwidth]{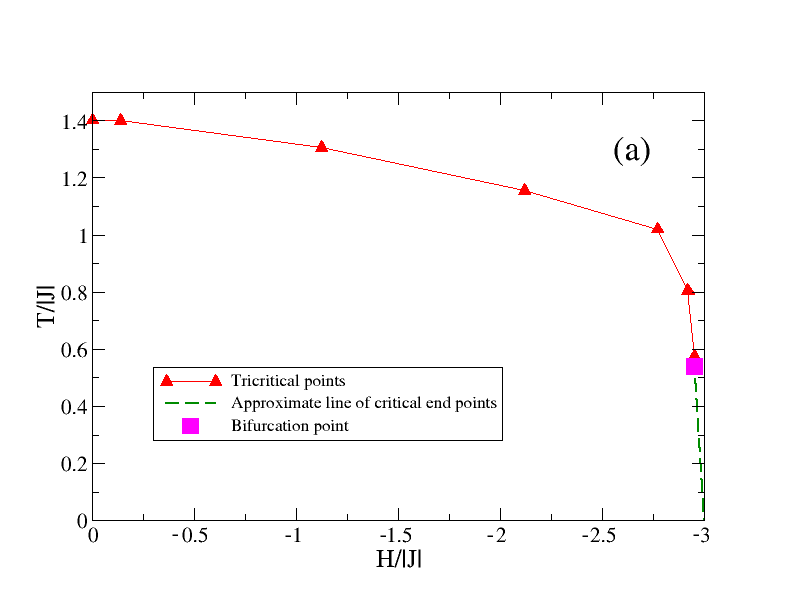}\hfill
    \includegraphics[width=.4\paperwidth]{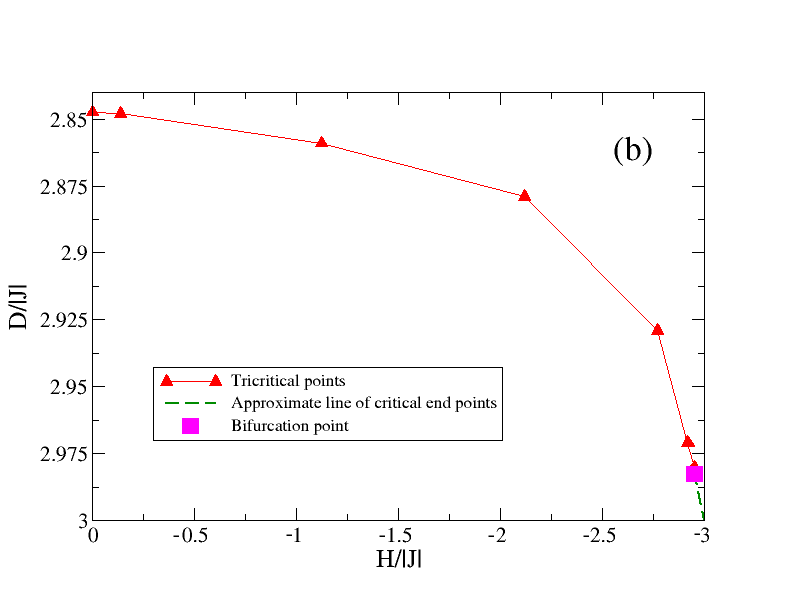}
  \end{flushleft}
    \caption{Projections of the lines of tricritical points and critical endpoints onto 
    (a) the $(-h,t)$ plane and 
    (b) the $(-h,-d)$ plane. The lines connecting the numerically calculated points are guides to 
    the eye. At this large scale, error bars would be smaller than the symbols. 
    }
    \label{fig:3CP}
  \end{figure}

 For the tricritical line at $h \ne 0$, we 
identify temperature regions where lines through nearby points identified as critical  and 
first-order, respectively, join smoothly together \cite{LAW84}. 
Within this range, the tricritical point  
is then identified by a scaling procedure analogous to the one shown for $h=0$ in 
Fig.~\ref{fig:3CPC}. The resulting, estimated tricritical line is shown in 
Fig.~\ref{fig:3CP}, projected onto the $(-h,t)$ and $(-h,-d)$ planes, respectively. 
The approximate locations of the bifurcation point and the line of critical endpoints shown in 
this figure are based on results discussed in Sec.~\ref{sec:Decomp}. 

\section{Detailed study of the bifurcation region}
\label{sec:Decomp}

As mentioned above, the two 
phase regions where decomposition of the tricritical line might be found
are restricted to approximately 
$|h| \in [2.94,3] $, $d \in [2.98,3]$, and $t < 0.57$.
Our approach is therefore to perform scans in $h$ or $t$ at fixed values of $d$ in this range.

\subsection{Second-order transitions}
\label{sec:second}

In this region of the phase diagram we also used susceptibility maxima and 
crossings of Binder cumulants for $L$ in the range $12,...,32$ to locate points on the 
surface of second-order transitions. The only difference is that we used scans in $h$ or $t$, 
instead of scans in $d$. 

\subsection{First-order transitions}
\label{sec:firstF}

  \subsubsection{Susceptibilities}
  \label{sec:susc1}

\begin{figure}[t]
    \centering
    \includegraphics[width=.5\textwidth]{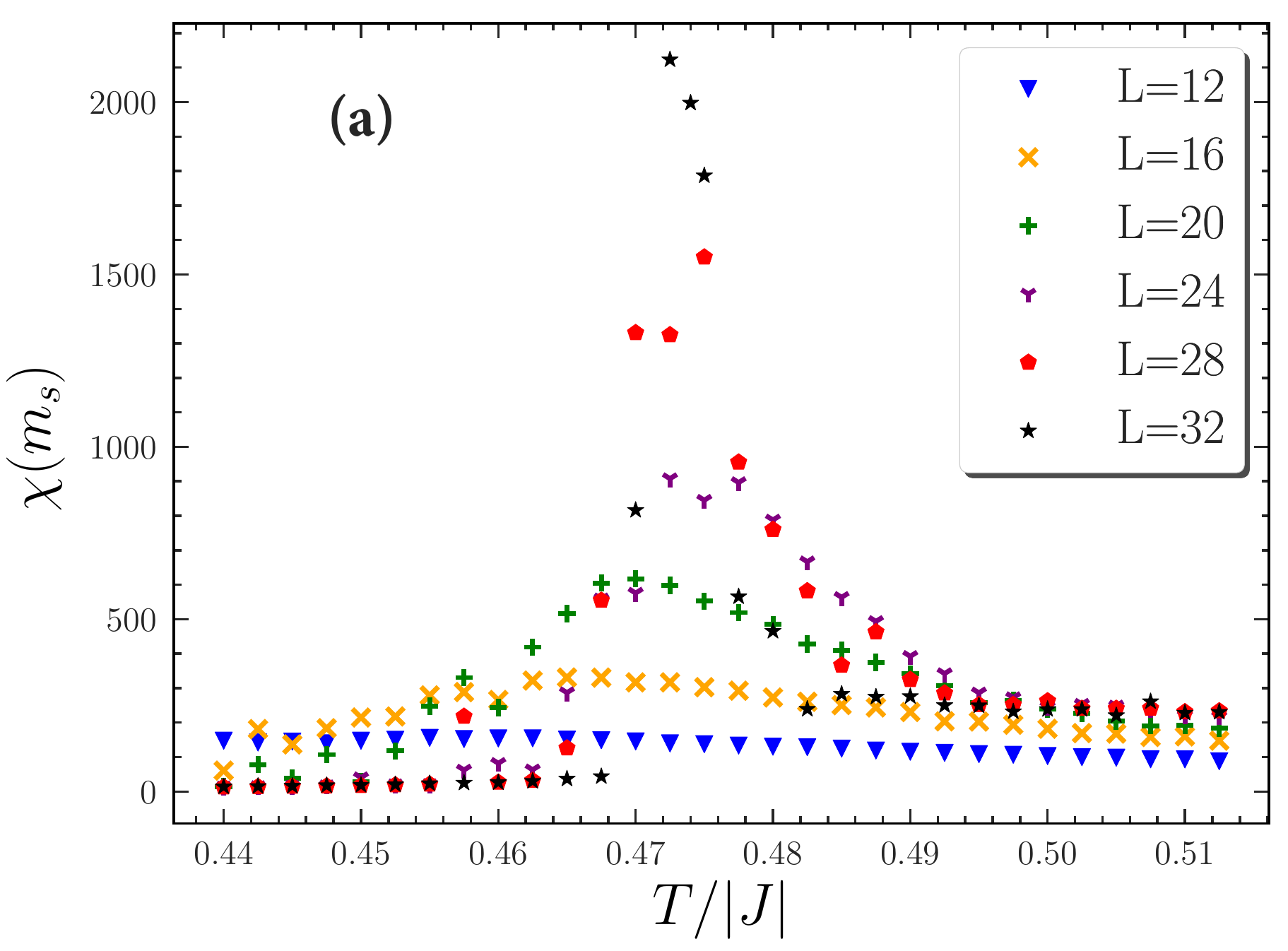}\hfill
   \includegraphics[width=.5\textwidth]{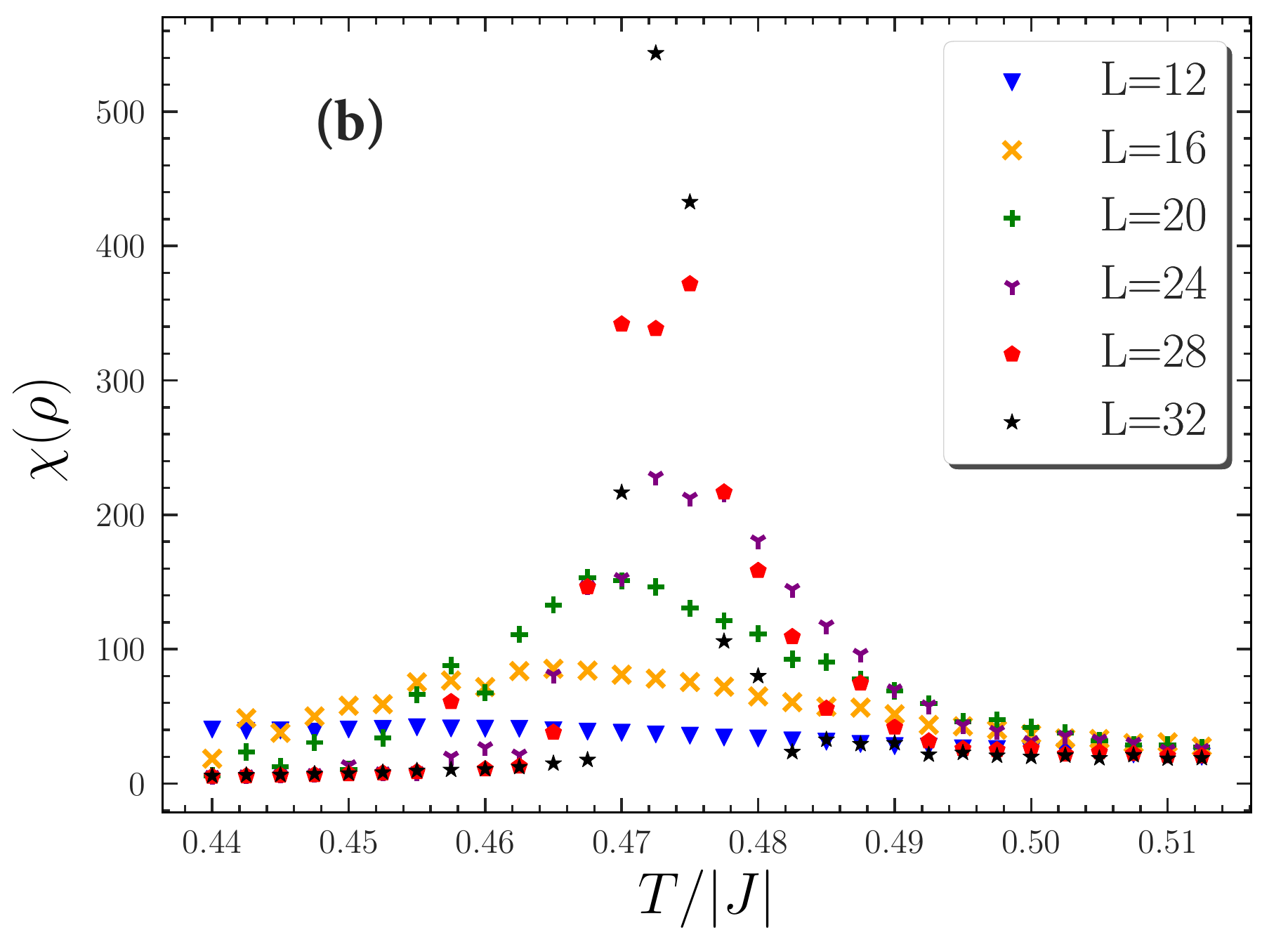}\\
    \includegraphics[width=.5\textwidth]{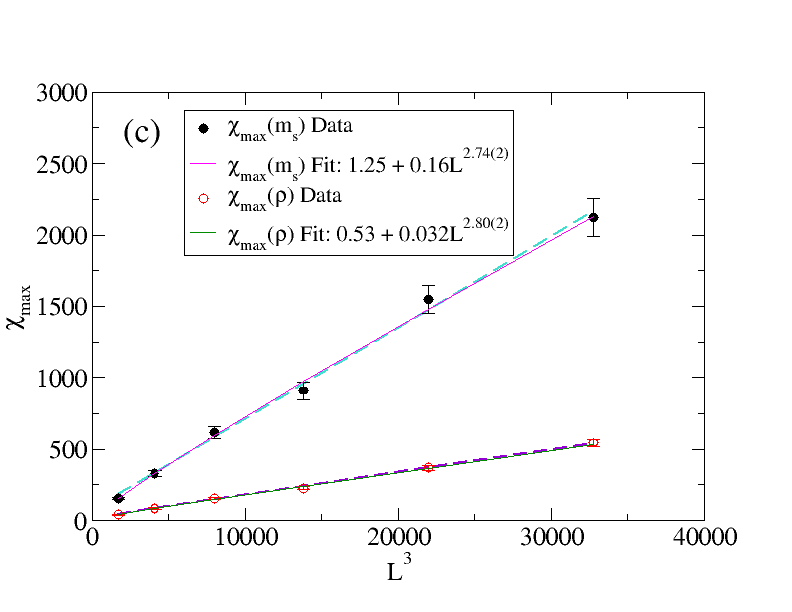}
    \caption{Details of the susceptibilities for $L = 12,\, ..., \, 32$, plotted vs $t$ across 
    the surface of first-order transitions for $d=2.9855$ and $h=-2.960$.
    Sharp maxima are seen at $t \approx 0.472$.
    {\bf (a)}: $\chi(m_{s})$. {\bf (b)}: $\chi( \rho)$. 
    {\bf (c)}: The maximum values of the two susceptibilities, plotted vs $L^3$. 
    The fitted scaling exponents are 
    2.74(2) for $\chi(m_{s})$ and 2.80(2) for $\chi( \rho)$, close to the expected 
    value of 3. 
    The dashed lines in the background are guides to the eye of form $a+bL^3$. 
    (The location of this phase point is shown as a green $\boldsymbol \times$  
in Figs.~\ref{fig:constant_dcuts} and \ref{fig:Projections}.) 
    }
    \label{fig:strong1st_order_susc}
  \end{figure}

As seen from their definition in Eq.~(\ref{eq:susc2}) and already noted in 
Sec.~\ref{sec:first}, the susceptibilities at a first-order transition 
should be asymptotically proportional to $L^3$. 
Plots of the susceptibilities for $m_s$ and $\rho$ vs $t$ 
at the same values of $d$ and $h$ as in 
Figs.~\ref{fig:free_energyl2p9855} and \ref{fig:strong1st_order_maxsuscbinder} below 
are presented in Fig.~\ref{fig:strong1st_order_susc} (a) and (b). As expected, the
susceptibility peaks grow larger and sharper as $L$ increases. 
  In Fig.~\ref{fig:strong1st_order_susc}(c) we show scaling plots of their maxima vs 
  $L^3$.  The scaling exponents of the fitted curves are 2.74(2) and 2.80(2), respectively, 
  close to the theoretically expected value of 3. 

\subsubsection{Order-parameter distributions and free energies}

\begin{figure}[tp]
    \centering
    \includegraphics[width=.45\textwidth]{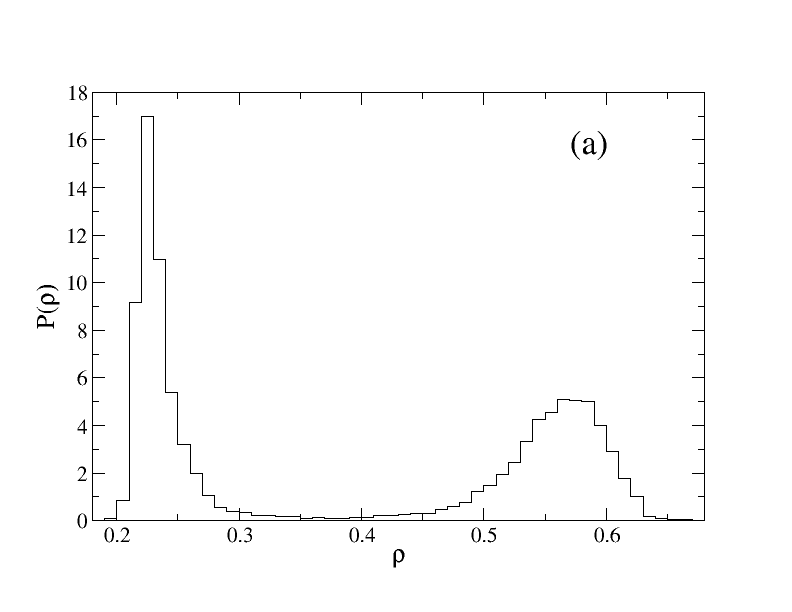}\hfill
    \includegraphics[width=.45\textwidth]{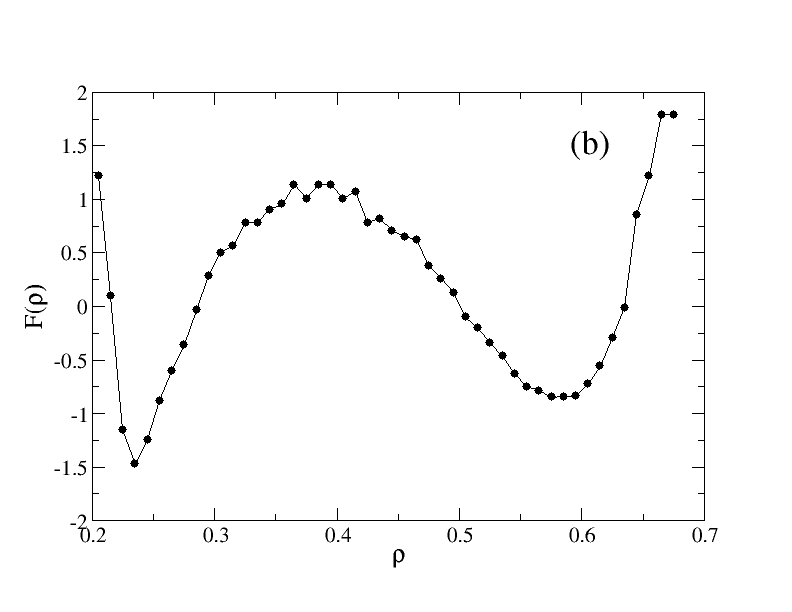}\\
    \caption{Normalized histogram with  $L=20$, representing $P(\rho)$ (a) and the 
    free energy $F(\rho)$ obtained from
    the histogram by Eq.~(\ref{eq:FO})  (b), representing a first-order transition   
    at $d=2.9805$, $h=-2.9394$, and $t=0.52$. 
    The areas under the histogram peaks are equal to within $\pm$2\%. 
    (The location of this phase point is shown as a red $\boldsymbol +$  in Figs.~\ref{fig:constant_dcuts} and \ref{fig:Projections}.) 
    }
    \label{fig:ms-densities}
  \end{figure}

\begin{figure}
    \centering
    \begin{tabular}{cc}
      \addheight{\includegraphics[width=.37\paperwidth]{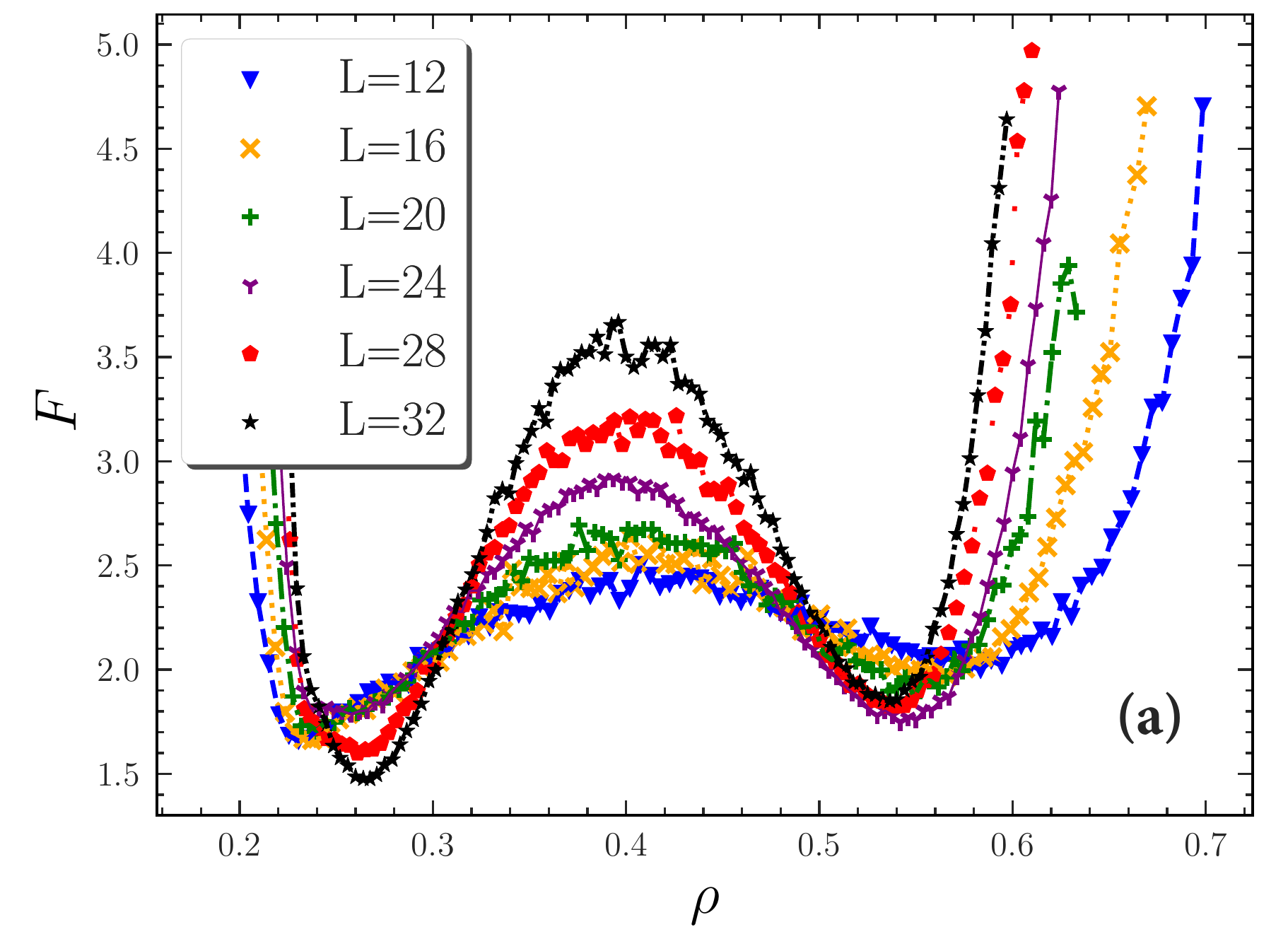}} &
      \addheight{\includegraphics[width=.39\paperwidth]{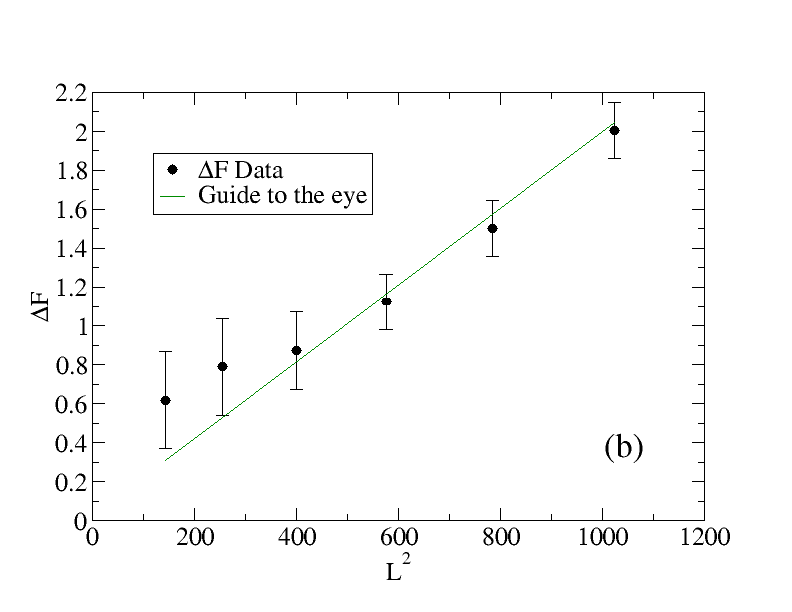}} \\
    \end{tabular}
    \caption{(a) Free energy $F$ vs density $\rho$ 
    at $d=2.9855$, $h=-2.9600$, and $t=0.475$ with $L=12$, ..., 32. 
    This phase point lies close to the susceptibility peaks in 
    Fig.~\ref{fig:strong1st_order_susc}, which are shown as 
   a green $\boldsymbol \times$ in Figs.~\ref{fig:constant_dcuts} and \ref{fig:Projections}.
    (b) Free- energy difference $\Delta F$  vs $L^2$. 
    Particularly for the larger systems, $\Delta F $ grows linearly with $L^2$ 
    as expected at a first-order phase transition.
    }
    \label{fig:free_energyl2p9855}
  \end{figure}

The probability distribution of an order parameter, $P({\mathcal O})$, has a very distinctive behavior that can be used to characterize the type of phase transition. As a first-order transition 
corresponds to a finite order-parameter discontinuity, $P({\mathcal O})$ has two well-defined peaks of equal area \cite{BIND84,BORG90,BORG92A,BORG92B}, 
as illustrated in Fig.~\ref{fig:ms-densities}(a).
Therefore, the free energy obtained from the probability distribution, 
\beq
F({\mathcal O}) = - T \ln P({\mathcal O}) \;,
\label{eq:FO}
\eeq
 has two valleys representing the distinct 
phases, separated by a local maximum representing the interface between them. 
(See Fig.~\ref{fig:ms-densities}(b).) 
The height of this maximum is given by 
\begin{equation}
\Delta F(L)=F(\rho_{max},L)-\cfrac{1}{2} [F(\rho _1,L)+F(\rho _2,L)] \;.
\label{eq:DeltaFr}
\end{equation}
Here, the relevant order parameter has been chosen as the density, $\rho$, 
and $\rho_1$, $\rho_2$, and $\rho_{max}$ refer to the two minima and the maximum 
in Fig.~\ref{fig:free_energyl2p9855}(a), respectively. 
For sufficiently large systems, the locations of the free-energy minima become independent 
of $L$, while their magnitudes continue to be size dependent.  
Since the free-energy maximum corresponds to 
an interface of dimension $d-1$, the asymptotic size dependence of 
$\Delta F$ is given by the finite-size scaling relation \cite{JLEE90, JLEE91},
\begin{equation}
\Delta F(L) \sim L^{d-1} \;.
\label{eq:DeltaFL}
\end{equation}
The scaling behavior of $\Delta F$ provides 
a sensitive method to locate a first-order phase transition. 
First, the transition point in the $(h,d,t)$ space for a given value of $L$ is identified from 
order-parameter histograms of several long simulation runs as the one that provides a pair of 
peaks of equal area. Repeating this procedure for a range of $L$, one can confirm the 
transition as first-order. 
Examples of the scaling plots corresponding to Eqs.~(\ref{eq:FO} - \ref{eq:DeltaFL}) for one particular point on the first-order surface 
are shown in Fig.~\ref{fig:free_energyl2p9855} (a) and (b), respectively. 
This method was also used, together with susceptibility scaling, to obtain 
the yellow points representing the first-order surface at higher 
temperatures in Fig.~\ref{fig:coasegbc}. 

\subsubsection{Binder Cumulant}
\label{sec:Binder}

\begin{figure}
    \begin{center}
      \addheight{\includegraphics[width=90mm]{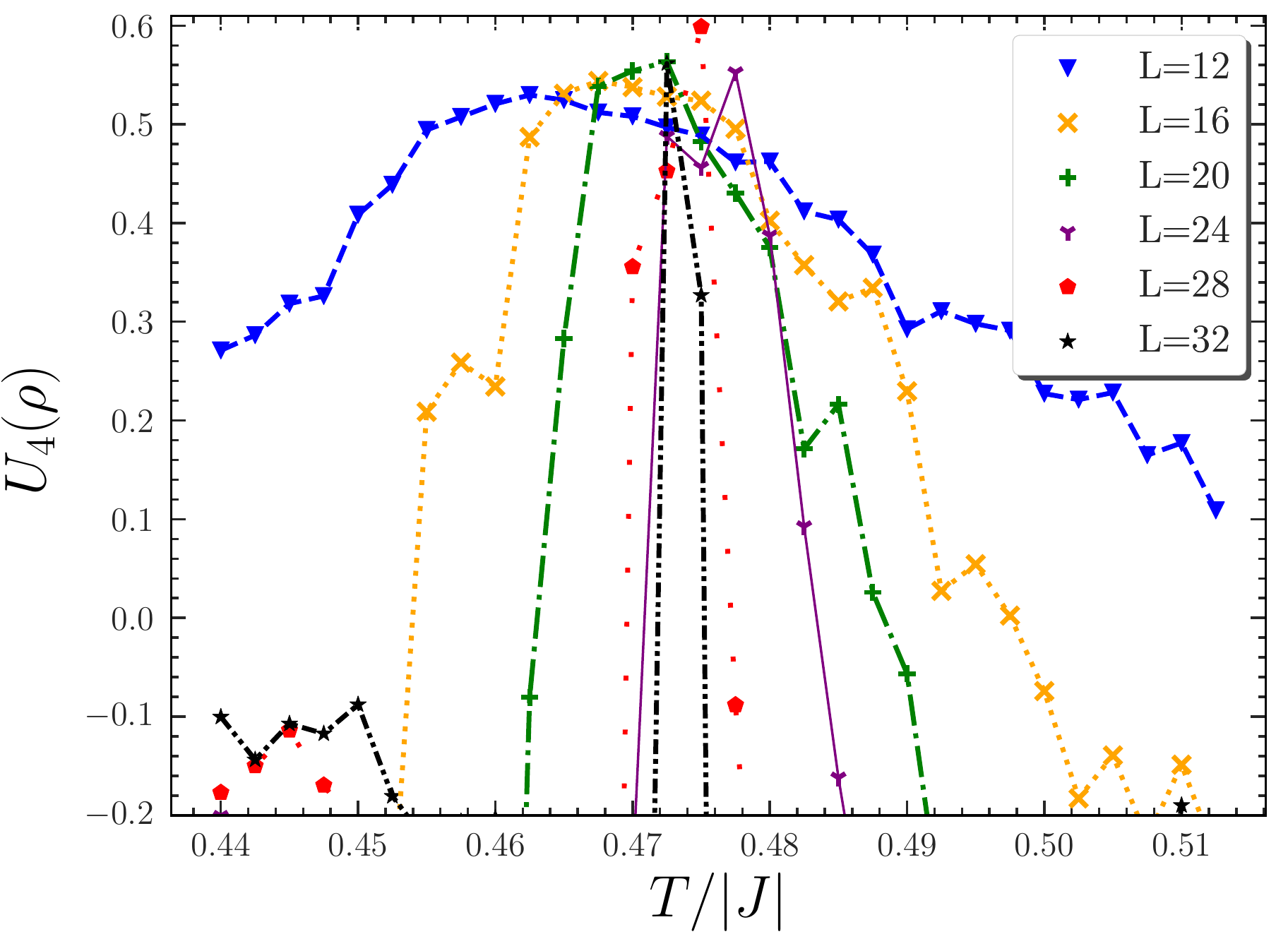}} 
    \caption{ Binder cumulant, $U_4$ vs $t$, at different system sizes at $d=2.9855,h=-2.960$. 
    It is negative around the sharper peaks that characterize the transition 
    temperature at $t = 0.474(2)$. 
     (The location of this phase point is shown as a green $\boldsymbol \times$ 
      in Figs.~\ref{fig:constant_dcuts} and \ref{fig:Projections}.) 
}
    \label{fig:strong1st_order_maxsuscbinder}
  \end{center}
  \end{figure}

The Binder cumulant [Eq.~(\ref{eq:Binder4})] 
also displays characteristic features at a first-order phase transition. 
Here, the cumulant shows a peak that becomes sharper at the transition as $L$ grows, 
and also reaches negative values on both sides, as shown in Fig. \ref{fig:strong1st_order_maxsuscbinder}. 
Both behaviors are indications of a first-order transition  \cite{VOLL93,TSAI98}. 
This method is particularly useful when no symmetries are available to help locate the transition 
\cite{CHAN15}. 
The peak positions for the largest values of $L$     
lie close to the susceptibility peaks shown in Fig.~\ref{fig:strong1st_order_susc}.

 \begin{figure}
    \begin{center}
      \hspace*{-10mm}
     \includegraphics[width=.6\textwidth]{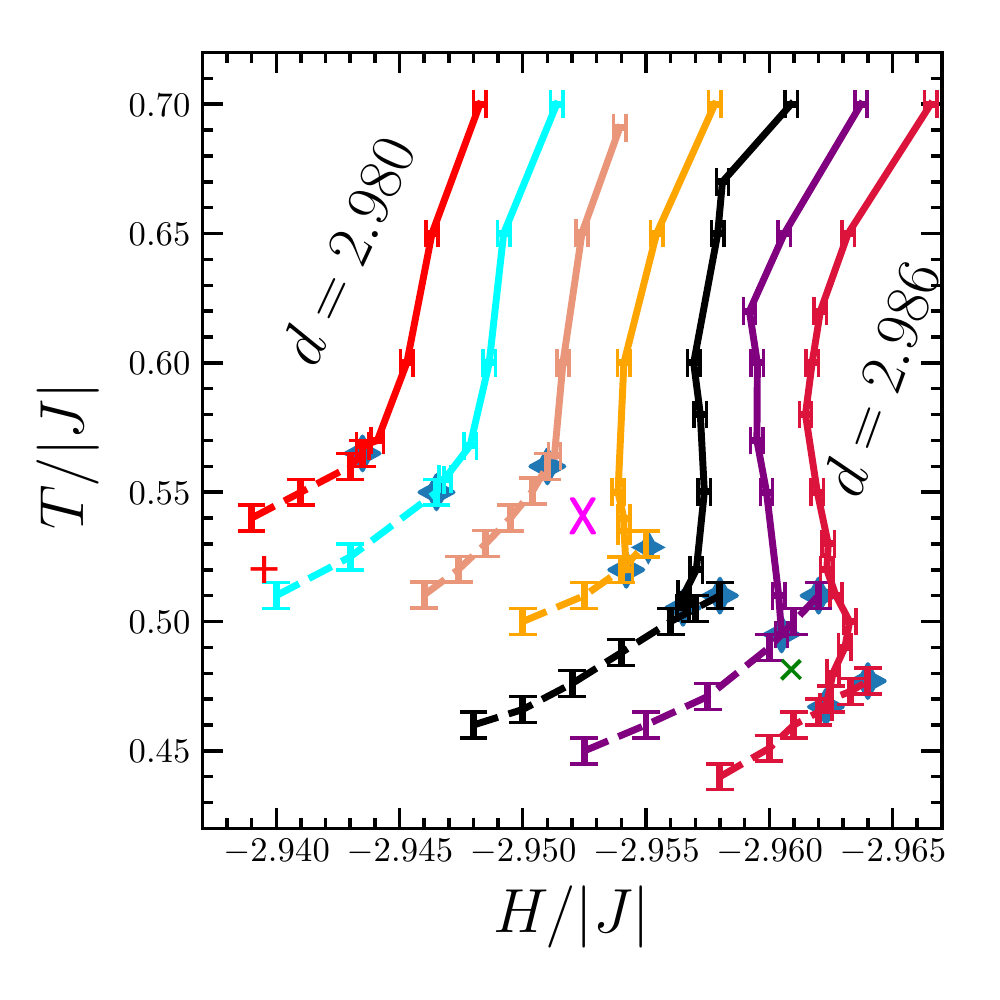}\hfill
      \end{center}
      \vspace*{-12mm}
    \caption{Sections of the phase diagram at constant values of 
    $d$, projected onto the $(-h,t)$ plane and 
    shown at intervals of $\Delta  d=0.001$ between 2.980 and 2.986. 
    The lines connecting the data points are guides to the eye. 
    Solid lines with error bars in the $h$ direction represent second-order transitions. 
    Dashed lines with error bars in the $t$ direction represent first-order transitions. 
    Projection of the estimated bifurcation point (see Fig.~\ref{fig:Projections}) 
    is shown as a large, magenta $\boldsymbol \times$. 
Blue stars in the background mark tricritical points, critical endpoints, and 
the critical points that terminate the lines of first-order transitions beyond the bifurcation point. 
    Projections of the first-order transition points considered in Fig.~\ref{fig:ms-densities} 
 and in Figs.~\ref{fig:strong1st_order_susc}, 
 \ref{fig:free_energyl2p9855}, and \ref{fig:strong1st_order_maxsuscbinder} 
    are shown as a red ${\boldsymbol +}$ and and green ${\boldsymbol \times}$, respectively.    
    See discussion in the text.
    }
    \label{fig:constant_dcuts}
  \end{figure}

 \begin{figure}
    \begin{center}
      \hspace*{-10mm}
     \includegraphics[width=.5\textwidth]{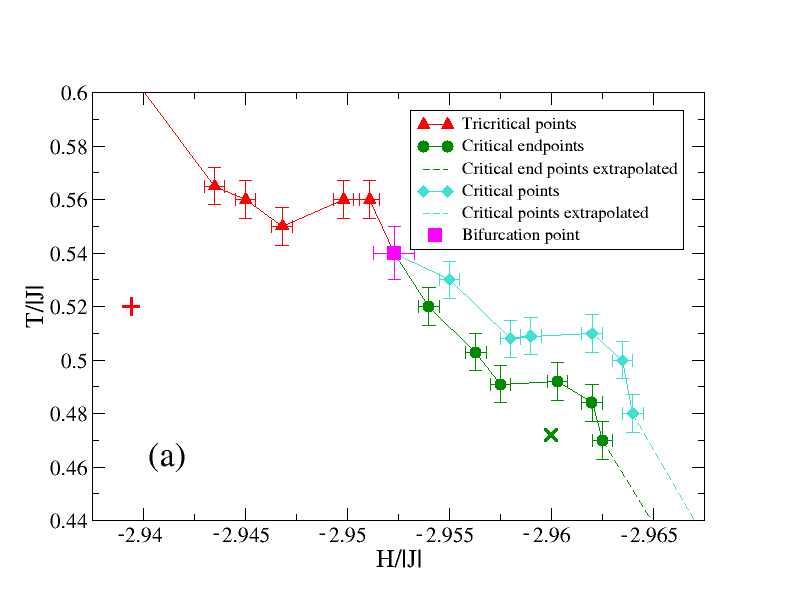}\hfill
     \includegraphics[width=.5\textwidth]{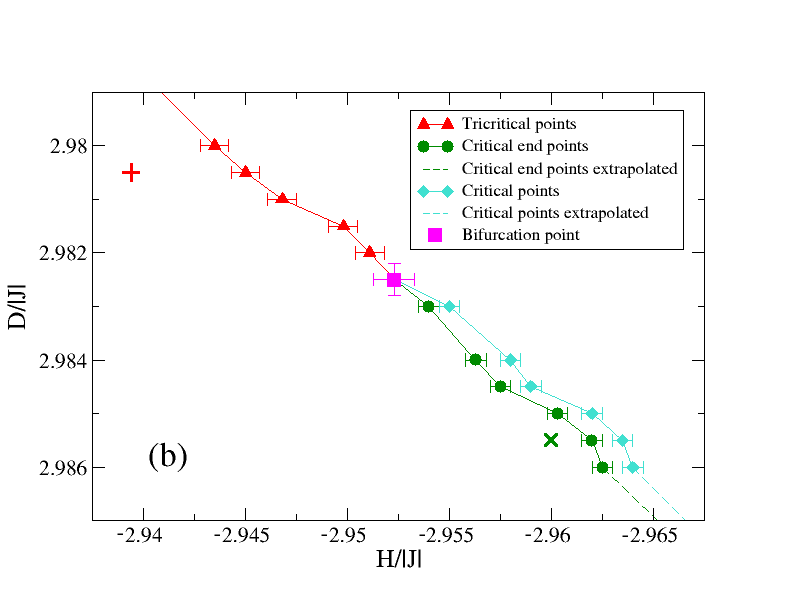}\\
      \end{center}
      \vspace*{-12mm}
    \caption{
    Lines of tricritical points (red), critical endpoints (green), and 
    critical points (turquoise). 
    The bifurcation point at $(h,d,t) = (-2.9523(1),2.9825(3),0.540(1))$ is 
    indicated as a magenta square. 
    Data and error bars extracted from twelve data sets at constant values of $d$, 
    seven of which are shown in Fig.~\ref{fig:constant_dcuts}. 
    This figure shows detail of the decomposition region of the phase diagram, 
    not visible on the scale of Fig.~\ref{fig:3CP}. 
    Phase points considered in Fig.~\ref{fig:ms-densities} and in 
    Figs.~\ref{fig:strong1st_order_susc} and \ref{fig:strong1st_order_maxsuscbinder} 
    are shown as a red ${\boldsymbol +}$ and and green ${\boldsymbol \times}$, respectively.
    (a): Projection onto the $(-h,t)$ plane. 
    (b): Projection onto the $(-h,-d)$ plane. 
    }
    \label{fig:Projections}
  \end{figure}

  \subsection{Decomposition of the tricritical line}
  \label{sec:decompX}
    
 In Fig.~\ref{fig:constant_dcuts} we present sections of the phase diagram for 
 constant values of $d$, slightly below 3. The first-order lines were obtained from histograms, 
 Binder cumulants, and susceptibilities, while the second-order lines were obtained from Binder 
cumulants and susceptibility maxima. 
Tricritical points at particular values of $d$ were identified as those where the first- and 
second-order lines join smoothly with a common slope. 
The order parameter selected was the density $\rho$. 
As $d$ increases, the tricritical points decompose into critical endpoints, where the first- and 
second-order lines meet at a finite angle, and 
critical points, separated by a line of first-order transitions. 
 [In the full $(d,h,t)$ space, this bifurcation would be seen as a point where a ``flap'' 
 of the surface of first-order transitions (yellow in Fig.~\ref{fig:coasegbc}) 
 ``dives below'' the second-order surface (blue in Fig.~\ref{fig:coasegbc}).]
 Based on the data in Fig.~\ref{fig:constant_dcuts} and additional simulations, 
 we estimate the bifurcation point to be at approximately 
 $(h,d,t) = ( -2.9523(1), 2.9825(3), 0.540(1))$. 
 This point is marked in the figure as a magenta $\boldsymbol \times$. 
 To within the mutual error bars, it agrees with the bifurcation value of $d$, 
 reported in \cite{KIME91A}. 
 
 The lines of tricritical points, critical endpoints, and critical points are shown in 
 Fig.~\ref{fig:Projections} as projections onto the $(-h,t)$ plane (a) and $(-h,-d)$ plane (b). 
 Extrapolations of the lines of critical endpoints and critical points are based on the 
 assumption that the two lines remain separated until they meet at $(h,d,t) = (-3,3,0)$. 
 The data for these plots were extracted from Fig.~\ref{fig:constant_dcuts} and additional 
 simulations at values of $d$, midway between the ones shown in that figure.

\begin{figure}
   \begin{center}
    \hspace*{-10mm}
    \includegraphics[width=.6\textwidth]{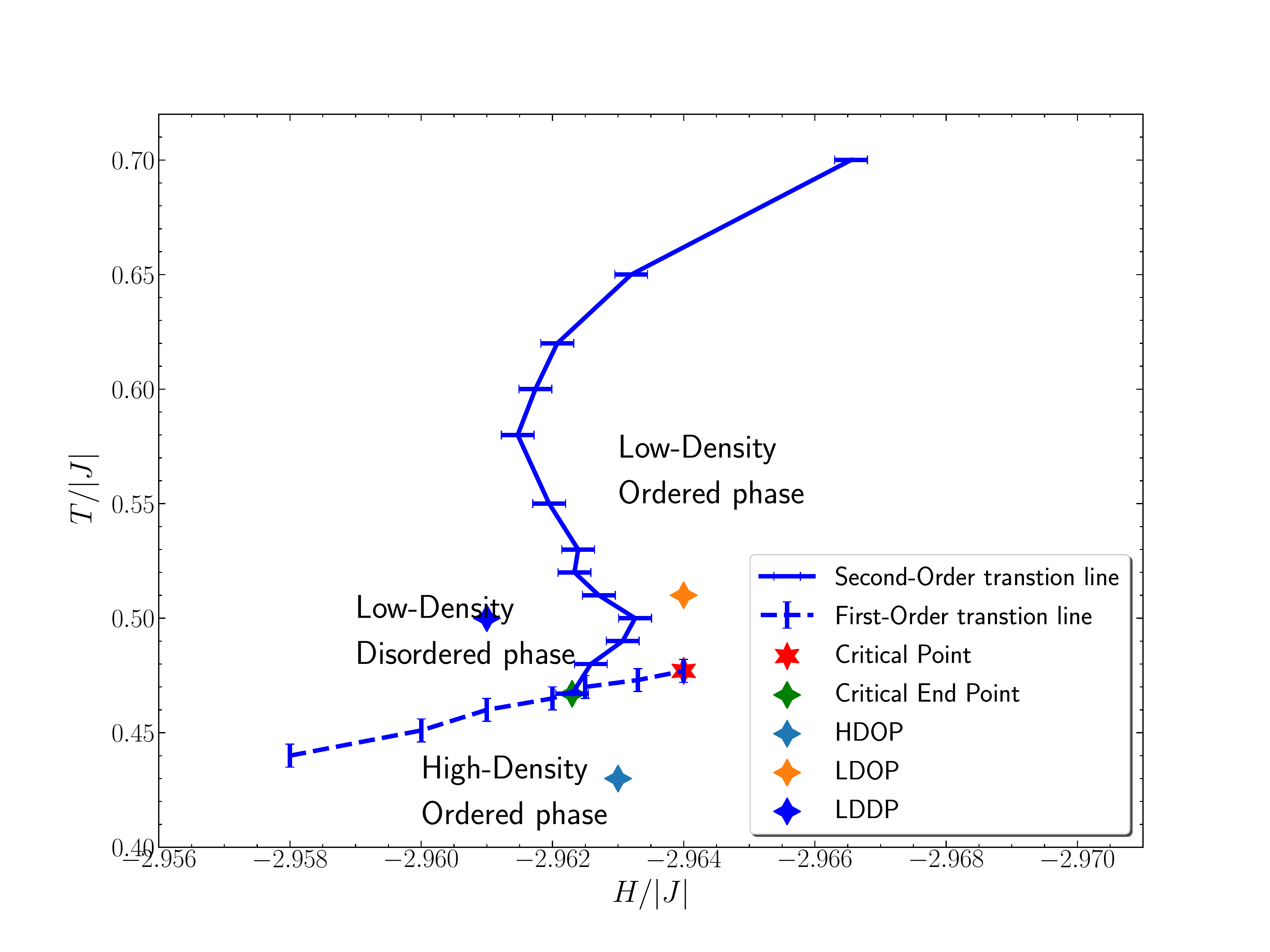}\hfill
  \end{center}
   \caption{Enlarged section of the phase diagram at $d=2.986$. The HDOP, LDOP, and LDDP phases are labeled, as well as the critical endpoint and the 
critical point that terminates the first-order line.
}
    \label{fig:d2p986}
  \end{figure}

   In Fig. \ref{fig:d2p986} we show in detail the phase diagram for  $d= 2.986$, where we identify 
  the lines of second-order and first-order transitions, as well as the critical endpoint, at which 
  they meet, and the critical point that terminates the first-order line inside the ordered-phase
   region. This result agrees well with
  Fig.~3 of Ref.~\cite{KIME91A}. Here we emphasize that it 
  permits three different phases: a low-density disordered phase (LDDP), which is 
  separated from a low-density ordered phase (LDOP) by the line of second-order transitions, 
  and from a high-density ordered phase (HDOP) by the line of first-order transitions. 
The LDOP and HDOP phases are separated by the section of the line of first-order transitions 
inside the ordered region, between the critical endpoint and the critical point.
 
 The degree of AFM ordering is represented by the staggered magnetization $m_s$, 
 and the density by $\rho$. 

 As the phase point is moved along the first-order line beyond the bifurcation, 
 order-parameter histograms 
 change from bimodal, with one peak representing a low-density ordered phase (LDOP) 
 and the other a 
 high-density ordered phase (HDOP), through 
 the critical point,
 to becoming a single peak 
 representing a single, ordered phase with values of 
 $m_s$ and $\rho$ that vary continuously with fields and temperature. 
 
\begin{figure}[htp]
    \centering
   \includegraphics[width=.7\textwidth]{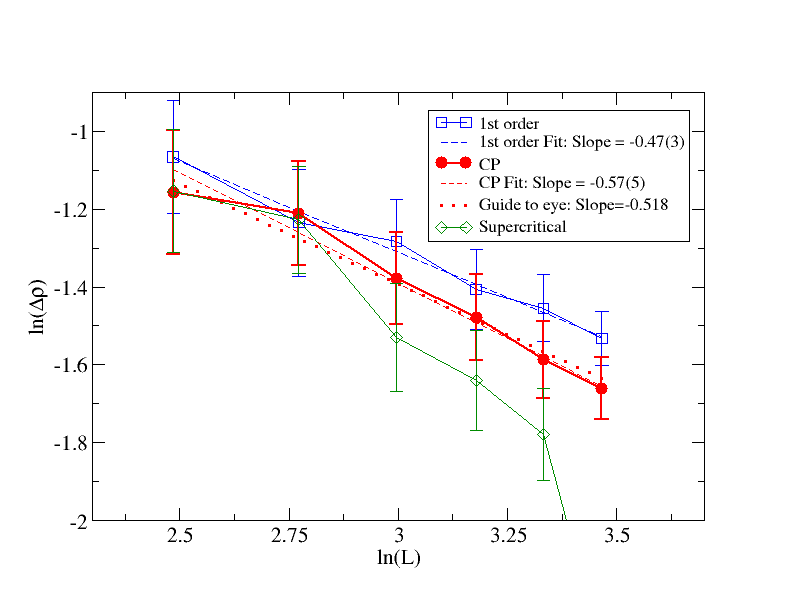}
    \caption
    {Log-log plot at $d = 2.9860$ of the distance, $\Delta \rho$, between the two peaks observed 
    in the density histogram for $L = 12, ..., 32$. Data for three phase points 
    in Fig.~\ref{fig:d2p986} are included. The red circles 
    correspond to our best estimate for the 
 critical point (CP) that terminates the first-order line  
    (red star in Fig.~\ref{fig:d2p986} at $h = -2.9640$ and $t = 0.4770$). 
    A weighted fit yields a slope of $-0.57(5)$, reasonably consistent with the expected value of 
    $- \beta / \nu \approx - 0.518$ for the 3D Ising universality class \cite{HASE10,RON17}. 
    The blue squares correspond to $h = -2.9633$ and $t= 0.4733$, marked with a vertical  
    bar on the order-order first-order line in Fig.~\ref{fig:d2p986}. The 
    magnitude of the fitted slope is smaller, 
    $-0.47(3)$, which we interpret as indicating an approach to an $L$-independent $\Delta \rho$ 
    in the large-$L$ limit. 
    The green diamonds correspond to $h = -2.645$ and $t = 0.4800$, on the 
    supercritical extension of the 
    first-order line into the uniform ordered phase. In this case, 
    $\Delta \rho$ decreases rapidly toward 0 with increasing $L$. 
}
    \label{fig:loglogdeltarho}
    \end{figure}

At the end of the first-order line inside the ordered-phase region, 
the distance between the two peaks 
 should decrease as $\Delta \rho \sim \ L^{-\beta/\nu}$ \cite{PRIV90B,HASE10,RON17}. 
 In Fig.~\ref{fig:loglogdeltarho} we show a log-log plot of $\Delta \rho$ vs $L$, at what we consider to be a good candidate for such a terminal critical point 
 at $d=2.986$, $h=-2.964$, and $t=0.477$ (red star in Fig.~\ref{fig:d2p986}). 
We estimate the slope of the log-log plot of $\Delta \rho$ as $-0.57 \pm 0.05$. 
Given the significantly different slopes for phase points at slightly higher and lower 
values of $h$, we consider this to be  
in reasonable agreement with the expected value for the 3D Ising universality class, 
$\beta / \nu \approx 0.518$ \cite{HASE10,RON17}.

\section{Phases}
\label{sec:Phas}

\subsection{Phase snapshots}
\label{sec:Snap}

\begin{figure}[htp]
    \centering
    \includegraphics[width=.55\textwidth]{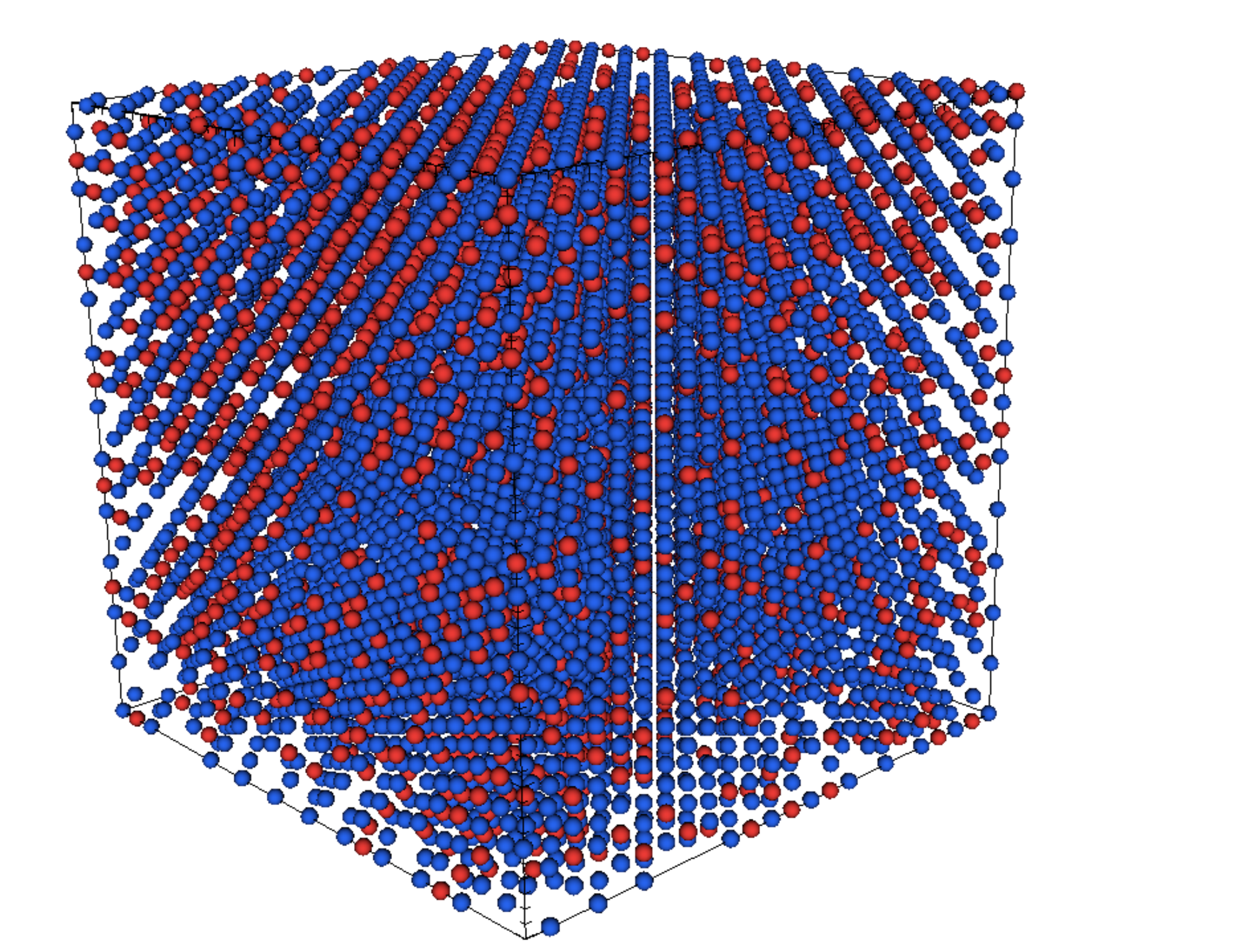}\hfill
    \caption{High-density ordered phase (HDOP), $d= 2.986, h= -2.963, t=0.43, L=24$. 
    This phase point is marked in Fig. \ref{fig:d2p986} by a light blue star. 
    Red points here represent $s=+1$, blue points represent $-1$, and vacant sites represent $0$. 
    As $h$ is negative, $s = -1$ (blue) is favored.
    This and the next two figures were created in VisIt \cite{VISIT12}.}
    \label{fig:hdphases}
  \end{figure}

\begin{figure}[htp]
    \centering
  \includegraphics[width=.55\textwidth]{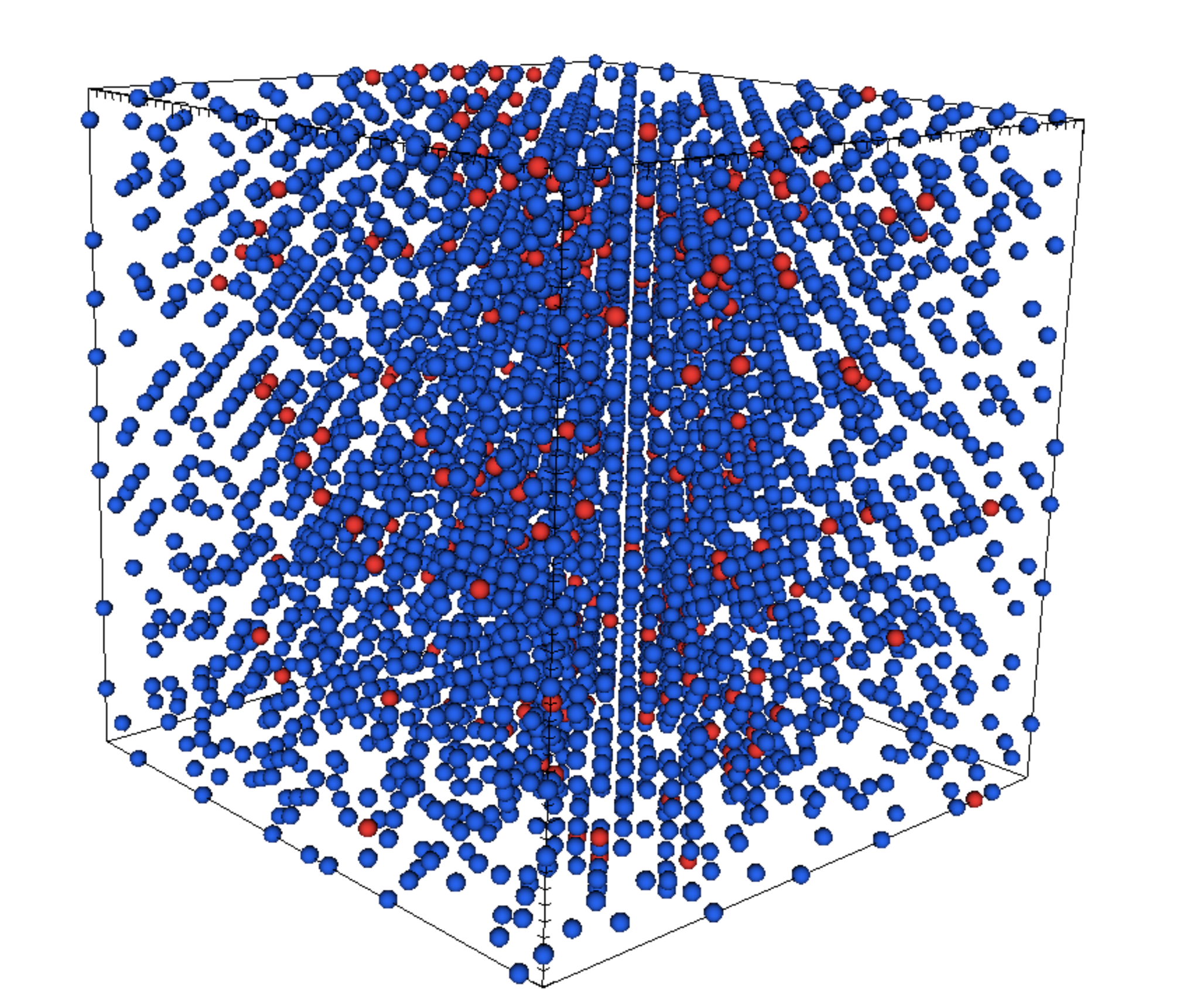}\hfill
  \caption{Low-density ordered phase (LDOP), $d= 2.986, h= -2.964, t=0.51, L=24$. This phase point is marked in Fig. \ref{fig:d2p986} by an orange star.}
  \label{fig:ld0phases}
  \end{figure}

\begin{figure}[htp]
    \centering
  \includegraphics[width=.55\textwidth]{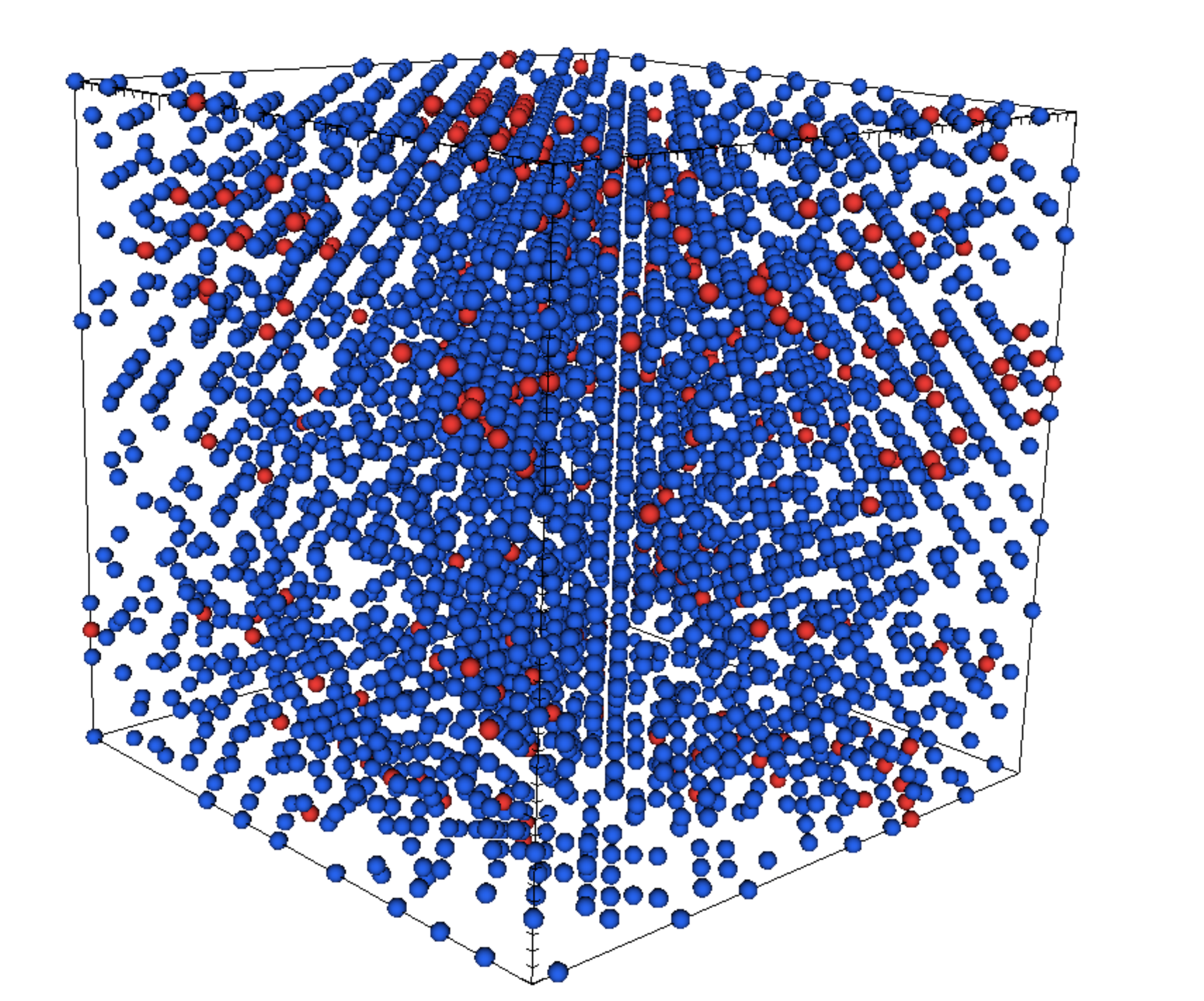}\\
  \caption{Low-density disordered phase (LDDP), $d= 2.986, h= -2.961, t=0.50, L=24$. This phase point is marked in Fig.~\ref{fig:d2p986} by a dark blue star.}
  \label{fig:lddphases}
  \end{figure}

Representative snapshots of the three phases, at phase points marked in Fig.~\ref{fig:d2p986}, 
 are shown in Figs.~\ref{fig:hdphases},  \ref{fig:ld0phases}, and \ref{fig:lddphases}, 
 for HDOP, LDOP, and LDDP, respectively.
The blue spheres represent $s = -1$, the red represent $s = +1$, while $s = 0$ (vacancies) are represented by empty sites. The HDOP (Fig.~\ref{fig:hdphases}) consists mostly of $s = -1$ alternating with $s = +1$, with about $40\%$ of vacancies scattered throughout. 
The vacancy density is clearly larger in the LD phases, Fig.~\ref{fig:ld0phases} 
and Fig.~\ref{fig:lddphases}, but it is relatively difficult to distinguish 
the LDOP and LDDP from the snapshots. 
Therefore, we next calculate the static structure factor for each phase.

\subsection{Static structure factors}
\label{sec:SQ}

In order to more clearly differentiate the phases, particularly the ordered and disordered 
low-density phases (LDOP and LDDP), we calculate their static structure factors. 
These are the Fourier transforms of the disconnected pair-correlation functions, 
\beq
  G(\vec{r}_2 - \vec{r}_1) \equiv  \langle s(\vec{r}_1) s(\vec{r}_2) \rangle \;,
\eeq
where $\vec{r}_i$ are the 3D lattice coordinates of the spin $s_i$. 
Structure factors are most easily evaluated as the square of the absolute value of the complex 
Fourier transform of the real-space spin configurations \cite{HASN13,KITTEL8}, 
\beq
S(k_x,k_y,k_z) = \left | \cfrac{1}{V} \sum_{x,y,z}   s(x,y,z) e^{- i (xk_x + yk_y + zk_z)  }  
\right |^2 \;.
\eeq
Here, $k_x =2\pi n_x /L = 2 \pi/ \lambda$ is the $x$ component of the wave vector, 
where $n_x$ is an integer that ranges from $0$ to $L-1$, and analogously for 
$k_y$ and $k_z$. The inverse volume, $1/V$, 
is the normalization factor, and $i$ is the imaginary unit. 
The $k_x$, $k_y$, and $k_z$ axes are plotted on $[0,2 \pi)$. 
The Fourier transform is normalized such that $S[\vec{k}=(0,0,0)] =m^2$. 
Also $S[\vec{k}=(\pi,\pi,\pi)] = m_s^2$ if the system is in a pure AFM configuration. 
A large and narrow peak represents a strongly AFM ordered system.

\begin{figure}
  \begin{tabular}{cc}
    (a) & (b) \\
    \addheight{\includegraphics[width=6cm,height=5.0cm]{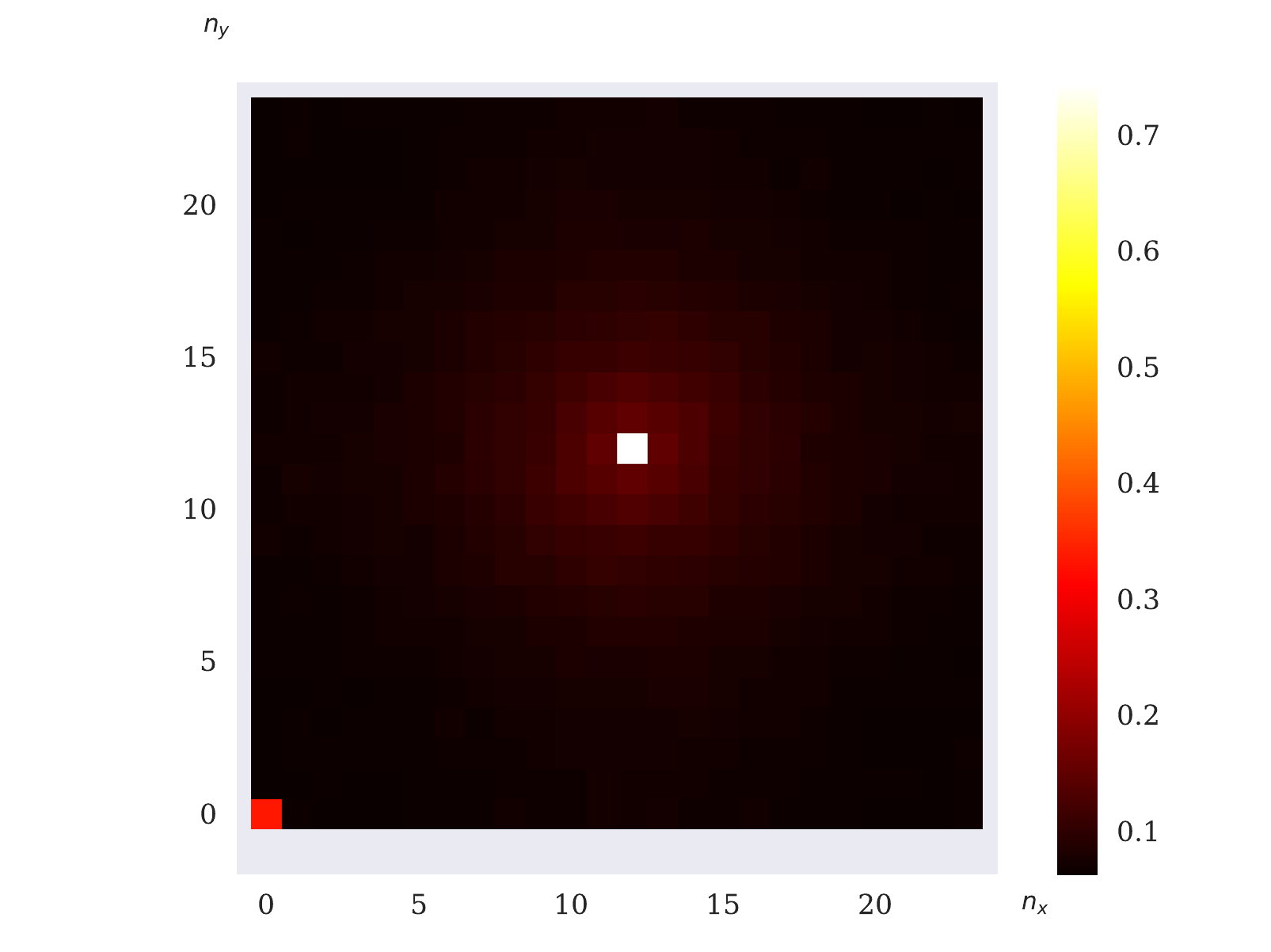}} &
    \addheight{\includegraphics[width=6cm,height=5.0cm]{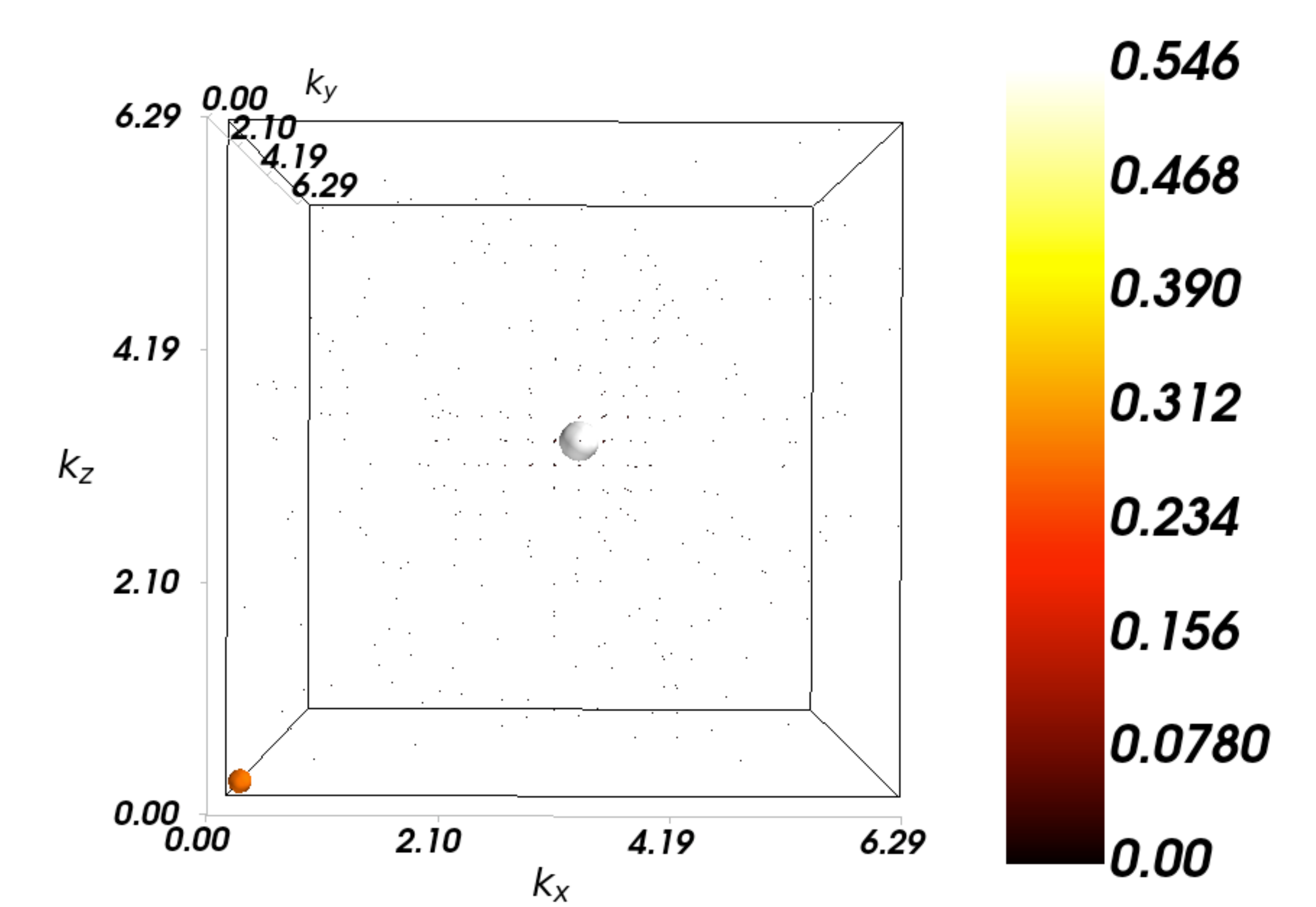}}\\
    \end{tabular}
  \caption{
  Structure factor averaged over 30 independent snapshots 
   at $d=2.986, t=0.43, h= -2.963, L=24$ (HDOP).
  (a) The 2D projected heat map, 
  $[ \tilde{S}(k_x,k_y) ]^{1/2} = \left[ \sum_{k_z} S(k_x,k_y,k_z) \right]^{1/2}$, and 
  (b) The 3D heat map, $[ S(k_x,k_y,k_z) ]^{1/2}$. 
  Note that the color scales are different in the two heat maps. 
  At $\vec{k} = (0,0,0)$, the magnetization magnitude is $|m| =0.276$, 
  and at $\vec{k} = (\pi,\pi,\pi)$, the magnitude of the staggered magnetization is 
  $|m_{s}| = 0.60$. 
    }
  \label{fig:structurehdop}
\end{figure}

\begin{figure}
  \begin{tabular}{cc}
    (a) & (b) \\
    \addheight{\includegraphics[width=6cm,height=5cm]{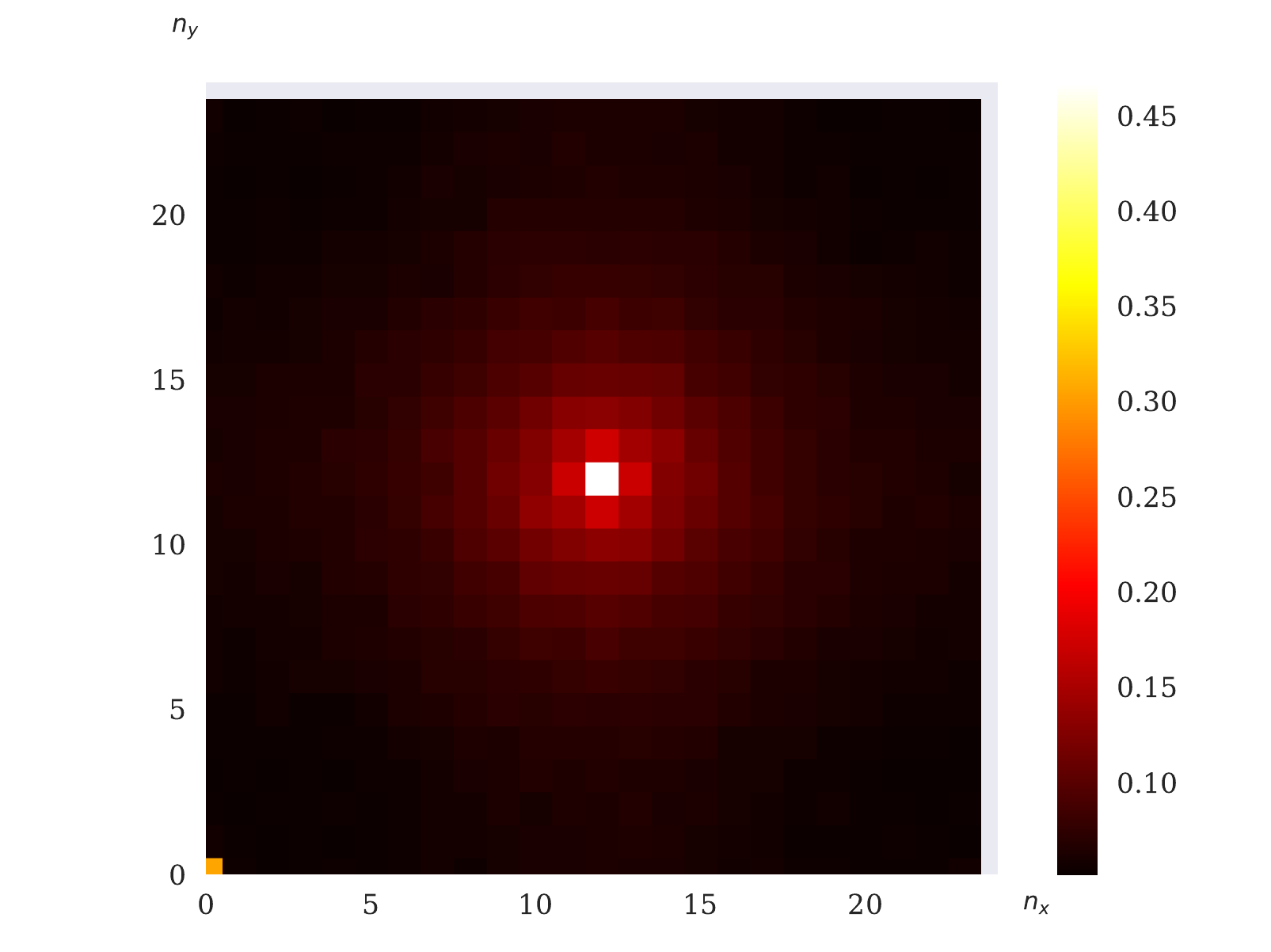}} &
    \addheight{\includegraphics[width=6cm,height=5cm]{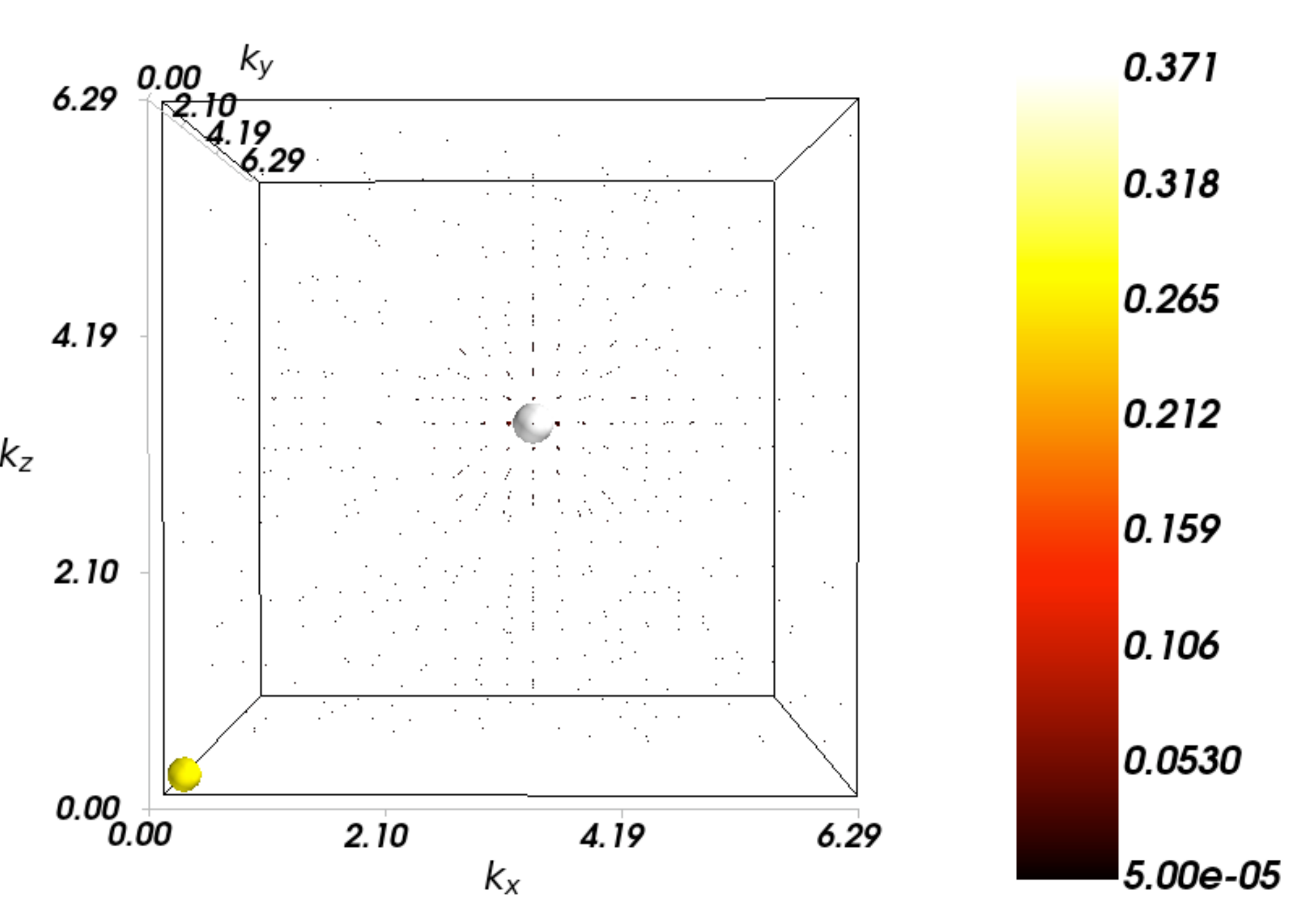}}\\
  \end{tabular}
  \caption{
   Structure factor averaged over 30 independent snapshots  at 
  $d=2.986, t=0.51, h= -2.964, L=24$ (LDOP).
  Otherwise as Fig.~\ref{fig:structurehdop}. 
  (a) The 2D projected heat map, and  (b) the 3D heat map. 
  At $\vec{k} = (0,0,0)$,  $|m|=0.253$, and at $\vec{k} = (\pi,\pi,\pi)$, $|m_{s}|= 0.302$. 
  The spread of intensity around the AFM peak indicates the 
  reduced ordering, compared to HDOP. 
  }
  \label{fig:structureldop}
 
\end{figure}

\begin{figure}
  \begin{tabular}{cc}
    (a) & (b) \\
    \addheight{\includegraphics[width=6cm,height=5cm]{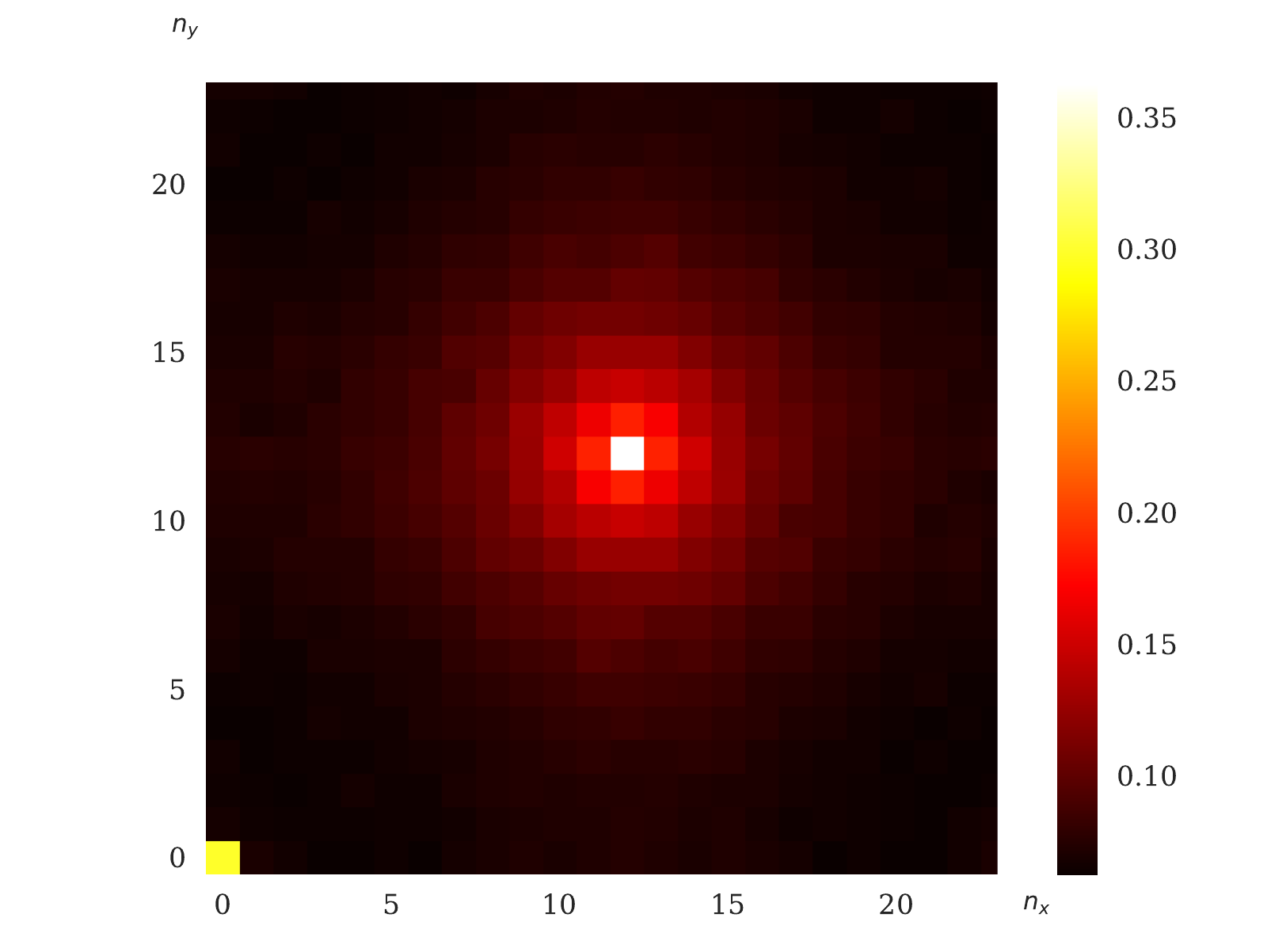}} &
\addheight{\includegraphics[width=6cm,height=5cm]{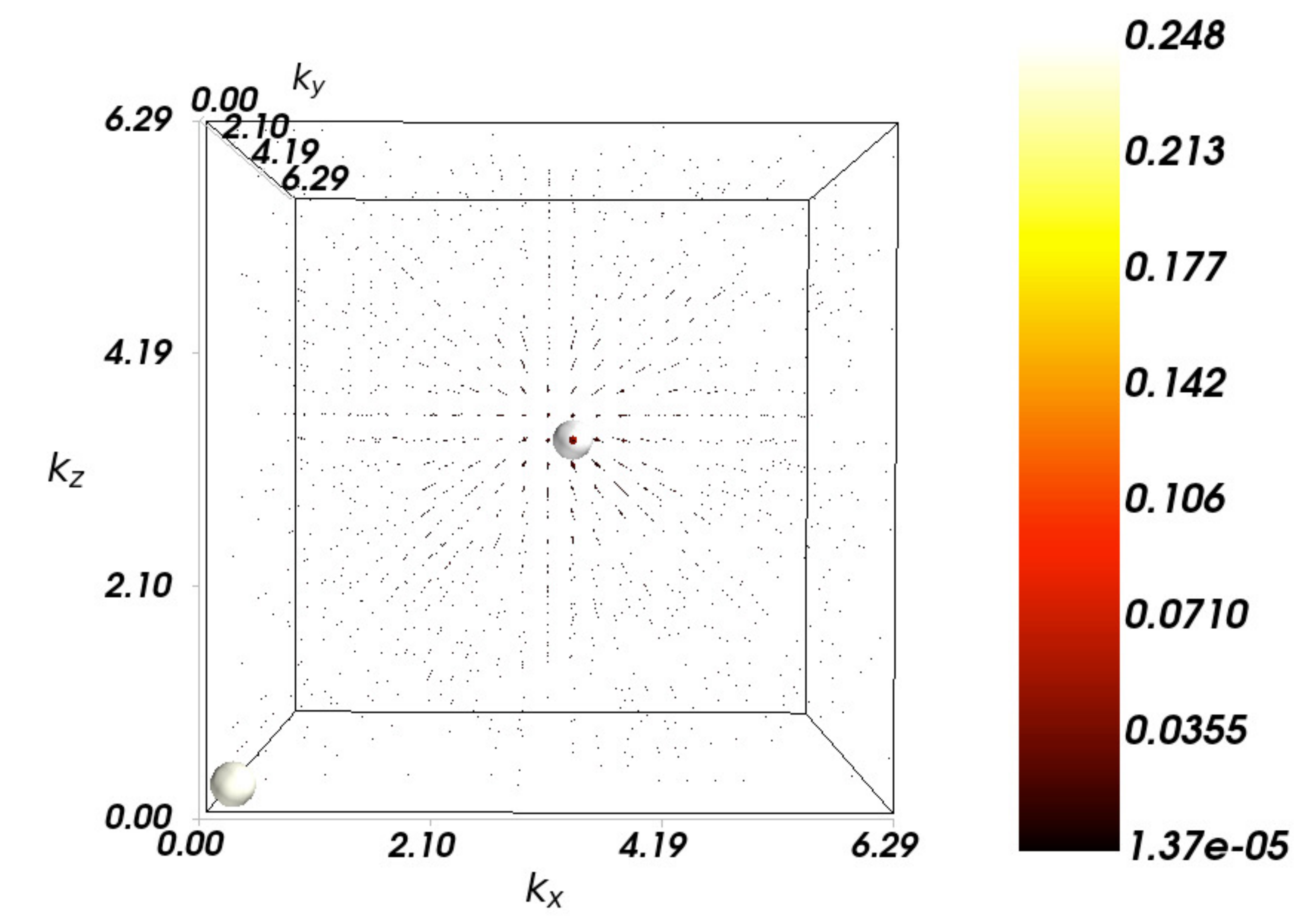}}\\
  \end{tabular}
  \caption{
  Structure factor averaged over 30 independent snapshots at 
  $d=2.986, t=0.50, h= -2.961, L=24$ (LDDP). 
  Otherwise as Fig.~\ref{fig:structurehdop}. 
  (a) The 2D projected heat map, and  (b) the 3D heat map. 
  At $\vec{k} = (0,0,0)$,  $|m| = 0.232$, 
  and at $\vec{k} = (\pi,\pi,\pi)$, $|m_{s}| = 0.248$. 
 The strong spread of intensity around the AFM peak indicates the 
  further reduced ordering, compared to LDOP. 
}
  \label{fig:structurelddp}
\end{figure}

The structure factors are shown in Figs. \ref{fig:structurehdop}, \ref{fig:structureldop}, and \ref{fig:structurelddp} for the HDOP, LDOP, and LDDP, respectively. The large sphere at 
$\vec{k}=(\pi,\pi,\pi)$ corresponds to AFM order. 
The sphere at the origin, $\vec{k}=(0,0,0)$, represents the magnetization  peak.

In the HDOP, the static structure factor has the largest AFM peak at $\vec{k}=(\pi,\pi,\pi)$ 
(Fig.~\ref{fig:structurehdop}), followed by the LDOP  (Fig.~\ref{fig:structureldop}), 
and then the LDDP (Fig.~\ref{fig:structurelddp}).
(Note that the color scales in the three figures are different.)
This is due to the fact that the HDOP has the largest amount of AFM ordering. The LDDP has the smallest and the most diffuse  AFM peak  of all, which indicates that other modes are important beside the AFM one. The clear difference between the structure factors of the 
LDDP and LDOP shows that these are, indeed, two different phases.

\begin{figure}
  \begin{center}
  \includegraphics[width=.47\textwidth]{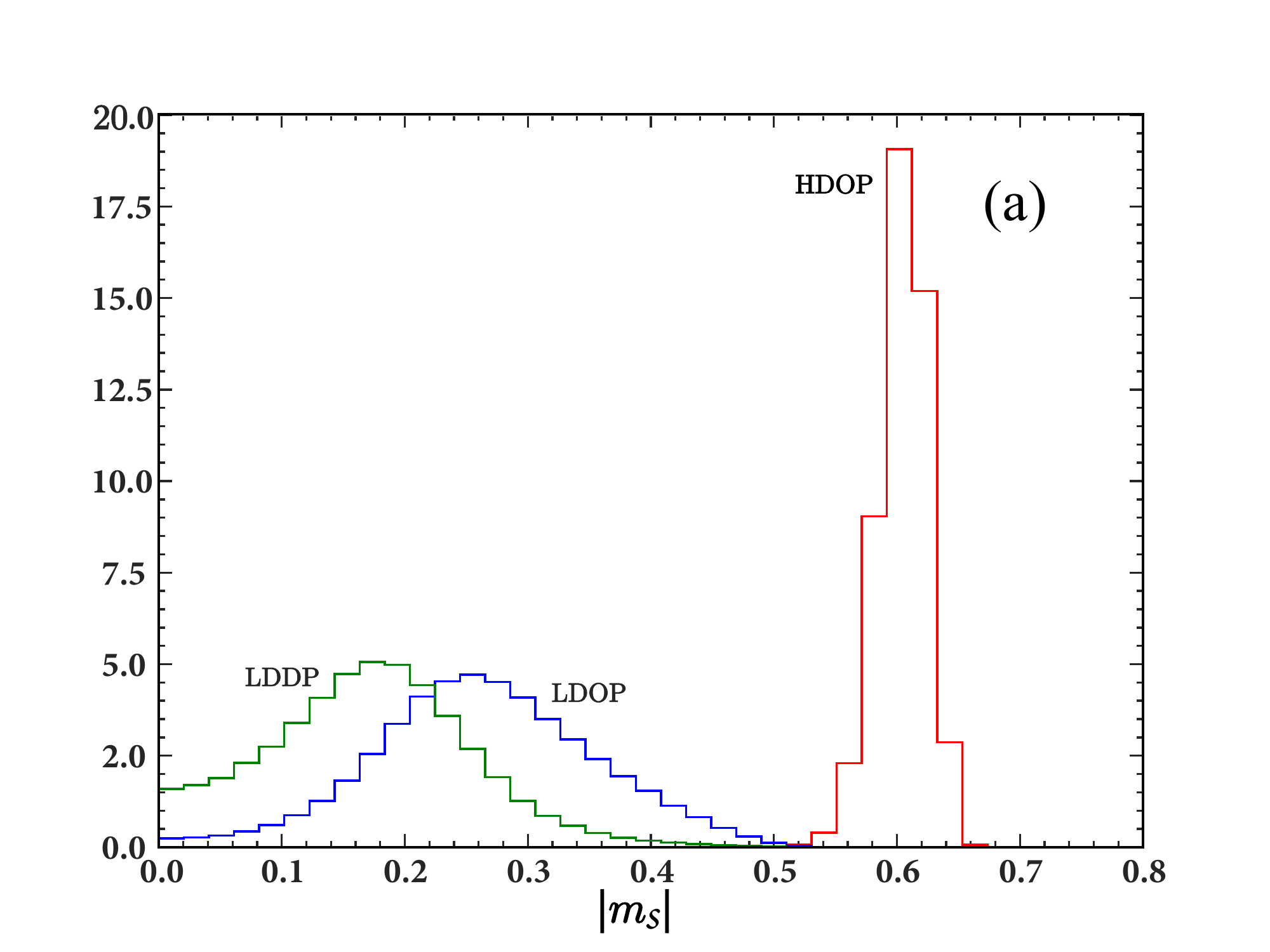} \hfill
  \includegraphics[width=.47\textwidth]{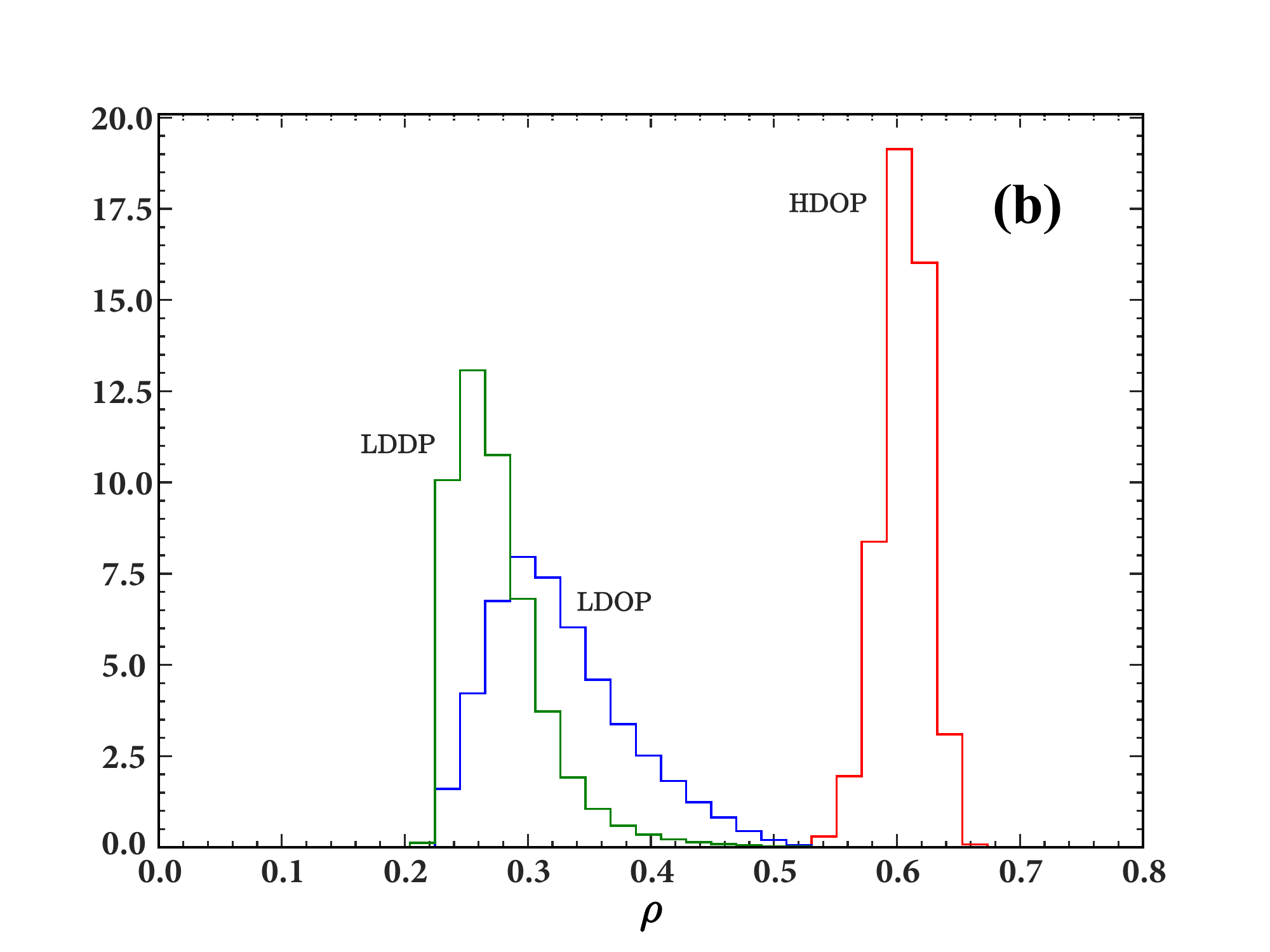} \\
  \end{center}
  \caption{Normalized histograms of the staggered magnetizations (a) and densities (b) corresponding to the 
  three phase points indicated in  Fig.\ref{fig:d2p986}. 
  Red corresponds to the light blue point in the HDOP, 
  blue to the orange point in the LDOP, 
  and green to the dark blue point in the LDDP.
  }
  \label{fig:hist_phases}
\end{figure}

Finally, we present the histograms corresponding to the staggered magnetization in 
Fig.~\ref{fig:hist_phases}(a) and the density in Fig.~\ref{fig:hist_phases}(b) for the three phase 
points selected in Fig.~\ref{fig:d2p986}, one in each phase. 
The histograms are clearly consistent with the phases indicated in the figure, and as we 
already noticed in the snapshots there is not a big difference between the low-density 
ordered and disordered phases.
Dividing the maximum-probability values of $|m_s|$ by those of $\rho$ in each phase, we obtain rough estimates of the proportions of the occupied sites in each phase that are AFM ordered: 
$0.60/0.60 = 1.00$ for HDOP, $0.25/0.30 = 0.83$ for LDOP, and  $0.16/0.26 = 0.62$ for LDDP.


  \section{Summary and Conclusions}
  \label{sec:Conc}

We have explored in detail the finite-temperature phase diagram of the 3D, 
antiferromagnetic Blume-Capel model on a simple cubic lattice, using Monte Carlo 
simulations and finite-size scaling analysis of susceptibilities, free energies, 
and Binder cumulants. The study consists of two major parts. 
 
First we considered, on a large scale, the overall phase diagram consisting of surfaces of 
second- and first-order phase transitions that join smoothly along a line of tricritical points 
 (Sec.~\ref{sec:large}). 
At $h=0$, we obtained the tricritical values of $d$ and $t$ 
in good agreement with previous results for the 
3D, ferromagnetic BC model \cite{DESE97,DENG04,ZIER15}, 
as well as the tricritical exponent ratios $\gamma/\nu$  in
excellent agreement with the theoretically expected values for the Ising universality class in 
three dimensions \cite{DENG04}.  

Second we considered, on much finer scales, the limited regions 
where decomposition of the tricritical line has been observed \cite{KIME91A}
(Secs.~\ref{sec:Decomp} and \ref{sec:Phas}). 
In Sec.~\ref{sec:Decomp}, surfaces of second- and first-order phase transitions  
were identified by finite-size scaling 
of data from scans in $h$ or $t$ on planes of constant $d$.  
The bifurcation point of the tricritical line was identified as the point where second- and 
first-order lines at constant $d$ 
changed from joining smoothly at the same angle (points on the tricritical line), 
to where the second-order line meets at a finite angle with a first-order line that continues into 
the ordered-phase region (critical endpoints), as seen in Fig.~\ref{fig:constant_dcuts}. 
The position of the bifurcation point in the $(h,d,t)$ space is in excellent agreement with 
the position reported in \cite{KIME91A}. 
Each first-order line that continues into the ordered-phase region in 
Fig.~\ref{fig:constant_dcuts} 
terminates at a critical point.
The lines of tricritical points, critical endpoints, and 
critical points are shown 
in projections onto the $(-h,t)$ and $(-h,-d)$ planes in Fig.~\ref{fig:Projections}. 
The surface bordered by the line of 
critical endpoints and the line of 
critical points is the ``flap'' of the surface of first-order transitions that continues 
into the ordered-phase ``volume,'' where it separates two different, ordered phases 
that become indistinguishable along the line of critical points.

Samples of these 
two ordered phases, as well as the disordered phase, 
were further investigated in Sec.~\ref{sec:Phas}. 
The clearest differentiation between the three phases is shown by the structure factors 
plotted in Figs.~\ref{fig:structurehdop}--\ref{fig:structurelddp}. 
As the phase becomes less strongly ordered, the antiferromagnetic maximum 
becomes increasingly diffuse. 

In this study we have constructed a comprehensive, multiscale picture of the 
topologically complex phase diagram of the Blume-Capel model on a simple cubic lattice. 
Three-state Ising or equivalent 
lattice-gas models with phase diagrams that involve intersecting surfaces of phase transitions  
are widely used to describe aspects of many  
physical and chemical systems. We therefore believe our results may provide inspiration 
for further applications of such models to real systems, introducing additional, local or 
long-range interactions and lattices of different dimensionality and symmetry. 
Beyond its interest as a study of static critical and multicritical properties in 
a multistate spin model, our work may also provide a starting point for dynamic studies of 
hysteresis and phase ordering at first-order transitions between differently ordered phases.

 \section*{Acknowledgments}
 G.B.\ would like to express her deep appreciation for support and hospitality at the 
 PoreLab and NJORD Centres of the Department of Physics at the University of Oslo, 
 and for the kindness of their personnel, making her stay very enjoyable and fruitful.
 
 We gratefully acknowledge useful comments on the manuscript by P.~Reis and M.~Moura. 
 
Work at the University of Oslo
was supported by the Research Council of Norway through the
Center of Excellence funding scheme, Project No. 262644.
Work at Florida State University was supported in part by the
U.S.\ Department of Energy Office of Science, Office of Nuclear
Physics under Award Number DE-FG02-92ER40750, and 
by U.S. National Science Foundation Grant No. DMR-1104829.


\begin{thebibliography}{43}%
\makeatletter
\providecommand \@ifxundefined [1]{%
 \@ifx{#1\undefined}
}%
\providecommand \@ifnum [1]{%
 \ifnum #1\expandafter \@firstoftwo
 \else \expandafter \@secondoftwo
 \fi
}%
\providecommand \@ifx [1]{%
 \ifx #1\expandafter \@firstoftwo
 \else \expandafter \@secondoftwo
 \fi
}%
\providecommand \natexlab [1]{#1}%
\providecommand \enquote  [1]{``#1''}%
\providecommand \bibnamefont  [1]{#1}%
\providecommand \bibfnamefont [1]{#1}%
\providecommand \citenamefont [1]{#1}%
\providecommand \href@noop [0]{\@secondoftwo}%
\providecommand \href [0]{\begingroup \@sanitize@url \@href}%
\providecommand \@href[1]{\@@startlink{#1}\@@href}%
\providecommand \@@href[1]{\endgroup#1\@@endlink}%
\providecommand \@sanitize@url [0]{\catcode `\\12\catcode `\$12\catcode
  `\&12\catcode `\#12\catcode `\^12\catcode `\_12\catcode `\%12\relax}%
\providecommand \@@startlink[1]{}%
\providecommand \@@endlink[0]{}%
\providecommand \url  [0]{\begingroup\@sanitize@url \@url }%
\providecommand \@url [1]{\endgroup\@href {#1}{\urlprefix }}%
\providecommand \urlprefix  [0]{URL }%
\providecommand \Eprint [0]{\href }%
\providecommand \doibase [0]{https://doi.org/}%
\providecommand \selectlanguage [0]{\@gobble}%
\providecommand \bibinfo  [0]{\@secondoftwo}%
\providecommand \bibfield  [0]{\@secondoftwo}%
\providecommand \translation [1]{[#1]}%
\providecommand \BibitemOpen [0]{}%
\providecommand \bibitemStop [0]{}%
\providecommand \bibitemNoStop [0]{.\EOS\space}%
\providecommand \EOS [0]{\spacefactor3000\relax}%
\providecommand \BibitemShut  [1]{\csname bibitem#1\endcsname}%
\let\auto@bib@innerbib\@empty
\bibitem [{\citenamefont {Blume}(1966)}]{BLUME66}%
  \BibitemOpen
  \bibfield  {author} {\bibinfo {author} {\bibfnamefont {M.}~\bibnamefont
  {Blume}},\ }\bibfield  {title} {\bibinfo {title} {Theory of the first-order
  magnetic phase change in {UO}$_2$},\ }\href@noop {} {\bibfield  {journal}
  {\bibinfo  {journal} {Phys.\ Rev.}\ }\textbf {\bibinfo {volume} {141}},\
  \bibinfo {pages} {517} (\bibinfo {year} {1966})}\BibitemShut {NoStop}%
\bibitem [{\citenamefont {Capel}(1966)}]{CAPE66}%
  \BibitemOpen
  \bibfield  {author} {\bibinfo {author} {\bibfnamefont {H.~W.}\ \bibnamefont
  {Capel}},\ }\bibfield  {title} {\bibinfo {title} {On the possibility of
  first-order phase transitions in {I}sing systems of triplet ions with
  zero-field splitting},\ }\href@noop {} {\bibfield  {journal} {\bibinfo
  {journal} {Physica}\ }\textbf {\bibinfo {volume} {32}},\ \bibinfo {pages}
  {966} (\bibinfo {year} {1966})}\BibitemShut {NoStop}%
\bibitem [{\citenamefont {Blume}\ \emph {et~al.}(1971)\citenamefont {Blume},
  \citenamefont {Emery},\ and\ \citenamefont {Griffiths}}]{BLUME71}%
  \BibitemOpen
  \bibfield  {author} {\bibinfo {author} {\bibfnamefont {M.}~\bibnamefont
  {Blume}}, \bibinfo {author} {\bibfnamefont {V.~J.}\ \bibnamefont {Emery}},\
  and\ \bibinfo {author} {\bibfnamefont {R.~B.}\ \bibnamefont {Griffiths}},\
  }\bibfield  {title} {\bibinfo {title} {Ising model for the $\lambda$
  transition and phase separation in {H}e$^3$-{H}e$^4$ mixtures},\ }\href@noop
  {} {\bibfield  {journal} {\bibinfo  {journal} {Phys.\ Rev.\ A}\ }\textbf
  {\bibinfo {volume} {4}},\ \bibinfo {pages} {1071} (\bibinfo {year}
  {1971})}\BibitemShut {NoStop}%
\bibitem [{\citenamefont {Collins}\ \emph {et~al.}(1989)\citenamefont
  {Collins}, \citenamefont {Sacramento}, \citenamefont {Rikvold},\ and\
  \citenamefont {Gunton}}]{COLL89}%
  \BibitemOpen
  \bibfield  {author} {\bibinfo {author} {\bibfnamefont {J.~B.}\ \bibnamefont
  {Collins}}, \bibinfo {author} {\bibfnamefont {P.}~\bibnamefont {Sacramento}},
  \bibinfo {author} {\bibfnamefont {P.~A.}\ \bibnamefont {Rikvold}},\ and\
  \bibinfo {author} {\bibfnamefont {J.~D.}\ \bibnamefont {Gunton}},\ }\bibfield
   {title} {\bibinfo {title} {Lateral interactions in catalyst poisoning},\
  }\href@noop {} {\bibfield  {journal} {\bibinfo  {journal} {Surf.\ Sci.}\
  }\textbf {\bibinfo {volume} {221}},\ \bibinfo {pages} {277} (\bibinfo {year}
  {1989})}\BibitemShut {NoStop}%
\bibitem [{\citenamefont {Zhang}\ \emph {et~al.}(1996)\citenamefont {Zhang},
  \citenamefont {Sung}, \citenamefont {Rikvold},\ and\ \citenamefont
  {Wieckowski}}]{JZHA95B}%
  \BibitemOpen
  \bibfield  {author} {\bibinfo {author} {\bibfnamefont {J.}~\bibnamefont
  {Zhang}}, \bibinfo {author} {\bibfnamefont {Y.-S.}\ \bibnamefont {Sung}},
  \bibinfo {author} {\bibfnamefont {P.~A.}\ \bibnamefont {Rikvold}},\ and\
  \bibinfo {author} {\bibfnamefont {A.}~\bibnamefont {Wieckowski}},\ }\bibfield
   {title} {\bibinfo {title} {Underpotential deposition of {C}u on {A}u(111) in
  sulfate-containing electrolytes: A theoretical and experimental study},\
  }\href@noop {} {\bibfield  {journal} {\bibinfo  {journal} {J.\ Chem.\ Phys.}\
  }\textbf {\bibinfo {volume} {104}},\ \bibinfo {pages} {5699} (\bibinfo {year}
  {1996})}\BibitemShut {NoStop}%
\bibitem [{\citenamefont {Benhouria}\ \emph {et~al.}(2018)\citenamefont
  {Benhouria}, \citenamefont {Essaoudi}, \citenamefont {Ainane}, \citenamefont
  {Ahuja},\ and\ \citenamefont {Dujardin}}]{BENH2018}%
  \BibitemOpen
  \bibfield  {author} {\bibinfo {author} {\bibfnamefont {Y.}~\bibnamefont
  {Benhouria}}, \bibinfo {author} {\bibfnamefont {I.}~\bibnamefont {Essaoudi}},
  \bibinfo {author} {\bibfnamefont {A.}~\bibnamefont {Ainane}}, \bibinfo
  {author} {\bibfnamefont {R.}~\bibnamefont {Ahuja}},\ and\ \bibinfo {author}
  {\bibfnamefont {F.}~\bibnamefont {Dujardin}},\ }\bibfield  {title} {\bibinfo
  {title} {Hysteresis loops and dielectric properties of a mixed spin
  {B}lume-{C}apel {I}sing ferroelectric nanowire},\ }\href@noop {} {\bibfield
  {journal} {\bibinfo  {journal} {Physica A}\ }\textbf {\bibinfo {volume}
  {506}},\ \bibinfo {pages} {499} (\bibinfo {year} {2018})}\BibitemShut
  {NoStop}%
\bibitem [{\citenamefont {Dudka}\ \emph {et~al.}(2016)\citenamefont {Dudka},
  \citenamefont {Kondrat}, \citenamefont {Kornyshev},\ and\ \citenamefont
  {Oshanin}}]{DUDKA2016}%
  \BibitemOpen
  \bibfield  {author} {\bibinfo {author} {\bibfnamefont {M.}~\bibnamefont
  {Dudka}}, \bibinfo {author} {\bibfnamefont {S.}~\bibnamefont {Kondrat}},
  \bibinfo {author} {\bibfnamefont {A.}~\bibnamefont {Kornyshev}},\ and\
  \bibinfo {author} {\bibfnamefont {G.}~\bibnamefont {Oshanin}},\ }\bibfield
  {title} {\bibinfo {title} {Phase behaviour and structure of a superionic
  liquid in nonpolarized nanoconfinement},\ }\href@noop {} {\bibfield
  {journal} {\bibinfo  {journal} {J.\ Phys.: Cond. Mat.}\ }\textbf {\bibinfo
  {volume} {28}},\ \bibinfo {pages} {464007} (\bibinfo {year}
  {2016})}\BibitemShut {NoStop}%
\bibitem [{\citenamefont {Dudka}\ \emph {et~al.}(2019)\citenamefont {Dudka},
  \citenamefont {Kondrat}, \citenamefont {B{'e}nichou}, \citenamefont
  {Kornyshev},\ and\ \citenamefont {Oshanin}}]{DUDKA2019}%
  \BibitemOpen
  \bibfield  {author} {\bibinfo {author} {\bibfnamefont {M.}~\bibnamefont
  {Dudka}}, \bibinfo {author} {\bibfnamefont {S.}~\bibnamefont {Kondrat}},
  \bibinfo {author} {\bibfnamefont {O.}~\bibnamefont {B{'e}nichou}}, \bibinfo
  {author} {\bibfnamefont {A.}~\bibnamefont {Kornyshev}},\ and\ \bibinfo
  {author} {\bibfnamefont {G.}~\bibnamefont {Oshanin}},\ }\bibfield  {title}
  {\bibinfo {title} {Superionic liquids in conducting nanoslits: A variety of
  phase transitions and ensuing charging behavior},\ }\href@noop {} {\bibfield
  {journal} {\bibinfo  {journal} {J.\ Chem.\ Phys.}\ }\textbf {\bibinfo
  {volume} {28}},\ \bibinfo {pages} {184105} (\bibinfo {year}
  {2019})}\BibitemShut {NoStop}%
\bibitem [{\citenamefont {Groda}\ \emph {et~al.}(2021)\citenamefont {Groda},
  \citenamefont {Dudka}, \citenamefont {Kornyshev}, \citenamefont {Oshanin},\
  and\ \citenamefont {Kondrat}}]{DUDKA2021}%
  \BibitemOpen
  \bibfield  {author} {\bibinfo {author} {\bibfnamefont {Y.}~\bibnamefont
  {Groda}}, \bibinfo {author} {\bibfnamefont {M.}~\bibnamefont {Dudka}},
  \bibinfo {author} {\bibfnamefont {A.}~\bibnamefont {Kornyshev}}, \bibinfo
  {author} {\bibfnamefont {G.}~\bibnamefont {Oshanin}},\ and\ \bibinfo {author}
  {\bibfnamefont {S.}~\bibnamefont {Kondrat}},\ }\bibfield  {title} {\bibinfo
  {title} {Superionic liquids in conducting nanoslits: Insights from theory and
  simulations},\ }\href@noop {} {\bibfield  {journal} {\bibinfo  {journal} {J.\
  Phys.\ Chem.\ C}\ }\textbf {\bibinfo {volume} {125}},\ \bibinfo {pages}
  {4968} (\bibinfo {year} {2021})}\BibitemShut {NoStop}%
\bibitem [{\citenamefont {Silva}\ and\ \citenamefont
  {Rikvold}(2019)}]{SILVA2019}%
  \BibitemOpen
  \bibfield  {author} {\bibinfo {author} {\bibfnamefont {D.}~\bibnamefont
  {Silva}}\ and\ \bibinfo {author} {\bibfnamefont {P.~A.}\ \bibnamefont
  {Rikvold}},\ }\bibfield  {title} {\bibinfo {title} {Complete catalog of
  ground-state diagrams for the general three-state lattice-gas model with
  nearest-neighbor interactions on a square lattice},\ }\href@noop {}
  {\bibfield  {journal} {\bibinfo  {journal} {Phys.\ Chem.\ Chem.\ Phys.}\
  }\textbf {\bibinfo {volume} {21}},\ \bibinfo {pages} {6216} (\bibinfo {year}
  {2019})}\BibitemShut {NoStop}%
\bibitem [{\citenamefont {Fefelov}\ \emph {et~al.}(2019)\citenamefont
  {Fefelov}, \citenamefont {Myshlyavtsev},\ and\ \citenamefont
  {Myshlyavtseva}}]{FEFE19}%
  \BibitemOpen
  \bibfield  {author} {\bibinfo {author} {\bibfnamefont {V.~F.}\ \bibnamefont
  {Fefelov}}, \bibinfo {author} {\bibfnamefont {A.~V.}\ \bibnamefont
  {Myshlyavtsev}},\ and\ \bibinfo {author} {\bibfnamefont {M.~D.}\ \bibnamefont
  {Myshlyavtseva}},\ }\bibfield  {title} {\bibinfo {title} {Complete analysis
  of phase diversity of the simplest adsorption model of a binary gas mixture
  for all sets of undirected interactions between nearest neighbors},\
  }\href@noop {} {\bibfield  {journal} {\bibinfo  {journal} {Adsorption}\
  }\textbf {\bibinfo {volume} {25}},\ \bibinfo {pages} {545} (\bibinfo {year}
  {2019})}\BibitemShut {NoStop}%
\bibitem [{\citenamefont {Lara}\ \emph {et~al.}(2022)\citenamefont
  {Lara}, \citenamefont {Correa},\ and\ \citenamefont
  {D{\'\i}az}}]{LARA22}%
  \BibitemOpen
  \bibfield  {author} {\bibinfo {author} {\bibfnamefont {D.~P.}~\bibnamefont
  {Lara}}, \bibinfo {author} {\bibfnamefont {H.}~\bibnamefont
  {Correa}},\ and\ \bibinfo {author} {\bibfnamefont {D.~S.}~\bibnamefont
  {D{\'\i}az}},\ }\bibfield  {title} {\bibinfo {title} {Antiferromagnetic
  {B}lume-{C}apel model of the disordered {F}e-{M}n-{A}l ternary system},\
  }\href@noop {} {\bibfield  {journal} {\bibinfo  {journal} {Phys.\ Rev.\ E}\
  }\textbf {\bibinfo {volume} {106}},\ \bibinfo {pages} {044114} (\bibinfo
  {year} {2022})}\BibitemShut {NoStop}%
\bibitem [{\citenamefont {Hasnaoui}\ and\ \citenamefont
  {Piekarewicz}(2013)}]{HASN13}%
  \BibitemOpen
  \bibfield  {author} {\bibinfo {author} {\bibfnamefont {K.~H.~O.}\
  \bibnamefont {Hasnaoui}}\ and\ \bibinfo {author} {\bibfnamefont
  {J.}~\bibnamefont {Piekarewicz}},\ }\bibfield  {title} {\bibinfo {title}
  {Charged {I}sing model of neutron star matter},\ }\href@noop {} {\bibfield
  {journal} {\bibinfo  {journal} {Phys.\ Rev.\ C}\ }\textbf {\bibinfo {volume}
  {88}},\ \bibinfo {pages} {025807} (\bibinfo {year} {2013})}\BibitemShut
  {NoStop}%
\bibitem [{\citenamefont {Lawrie}\ and\ \citenamefont {Sarbach}(1984)}]{LAW84}%
  \BibitemOpen
  \bibfield  {author} {\bibinfo {author} {\bibfnamefont {I.~D.}\ \bibnamefont
  {Lawrie}}\ and\ \bibinfo {author} {\bibfnamefont {S.}~\bibnamefont
  {Sarbach}},\ }\bibfield  {title} {\bibinfo {title} {Theory of tricritical
  points},\ }in\ \href@noop {} {\emph {\bibinfo {booktitle} {Phase Transitions
  and Critical Phenomena, Vol.~9}}},\ \bibinfo {editor} {edited by\ \bibinfo
  {editor} {\bibfnamefont {C.}~\bibnamefont {Domb}}\ and\ \bibinfo {editor}
  {\bibfnamefont {J.~L.}\ \bibnamefont {Lebowitz}}}\ (\bibinfo  {publisher}
  {Academic Press},\ \bibinfo {address} {London},\ \bibinfo {year} {1984})\
  pp.\ \bibinfo {pages} {1--161}\BibitemShut {NoStop}%
\bibitem [{\citenamefont {Collins}\ \emph {et~al.}(1988)\citenamefont
  {Collins}, \citenamefont {Rikvold},\ and\ \citenamefont
  {Gawlinski}}]{COLL88}%
  \BibitemOpen
  \bibfield  {author} {\bibinfo {author} {\bibfnamefont {J.~B.}\ \bibnamefont
  {Collins}}, \bibinfo {author} {\bibfnamefont {P.~A.}\ \bibnamefont
  {Rikvold}},\ and\ \bibinfo {author} {\bibfnamefont {E.~T.}\ \bibnamefont
  {Gawlinski}},\ }\bibfield  {title} {\bibinfo {title} {Finite-size scaling
  analysis of the {$S$=1} {I}sing model on the triangular lattice},\
  }\href@noop {} {\bibfield  {journal} {\bibinfo  {journal} {Phys.\ Rev.\ B}\
  }\textbf {\bibinfo {volume} {38}},\ \bibinfo {pages} {6741} (\bibinfo {year}
  {1988})}\BibitemShut {NoStop}%
\bibitem [{\citenamefont {Wilding}(1997)}]{WILD97}%
  \BibitemOpen
  \bibfield  {author} {\bibinfo {author} {\bibfnamefont {N.~B.}\ \bibnamefont
  {Wilding}},\ }\bibfield  {title} {\bibinfo {title} {Coexistence curve
  singularities at critical end points},\ }\href@noop {} {\bibfield  {journal}
  {\bibinfo  {journal} {Phys.\ Rev.\ Lett.}\ }\textbf {\bibinfo {volume}
  {78}},\ \bibinfo {pages} {1488} (\bibinfo {year} {1997})}\BibitemShut
  {NoStop}%
\bibitem [{\citenamefont {Wang}\ and\ \citenamefont
  {Rauchwarger}(1976)}]{WANG1976}%
  \BibitemOpen
  \bibfield  {author} {\bibinfo {author} {\bibfnamefont {Y.-L.}\ \bibnamefont
  {Wang}}\ and\ \bibinfo {author} {\bibfnamefont {K.}~\bibnamefont
  {Rauchwarger}},\ }\bibfield  {title} {\bibinfo {title} {Multicritical
  behavior in an {I}sing antiferromagnet with zero field splitting},\
  }\href@noop {} {\bibfield  {journal} {\bibinfo  {journal} {Phys.\ Lett.\ A}\
  }\textbf {\bibinfo {volume} {59}},\ \bibinfo {pages} {73} (\bibinfo {year}
  {1976})}\BibitemShut {NoStop}%
\bibitem [{\citenamefont {Wang}\ and\ \citenamefont {Kimel}(1991)}]{KIME91A}%
  \BibitemOpen
  \bibfield  {author} {\bibinfo {author} {\bibfnamefont {Y.-L.}\ \bibnamefont
  {Wang}}\ and\ \bibinfo {author} {\bibfnamefont {J.~D.}\ \bibnamefont
  {Kimel}},\ }\bibfield  {title} {\bibinfo {title} {Multicritical behavior in
  the antiferromagnetic {B}lume-{C}apel model},\ }\href@noop {} {\bibfield
  {journal} {\bibinfo  {journal} {J.\ Appl.\ Phys.}\ }\textbf {\bibinfo
  {volume} {69}},\ \bibinfo {pages} {6176} (\bibinfo {year}
  {1991})}\BibitemShut {NoStop}%
\bibitem [{\citenamefont {Kimel}\ \emph {et~al.}(1992)\citenamefont {Kimel},
  \citenamefont {Rikvold},\ and\ \citenamefont {Wang}}]{KIME91}%
  \BibitemOpen
  \bibfield  {author} {\bibinfo {author} {\bibfnamefont {J.~D.}\ \bibnamefont
  {Kimel}}, \bibinfo {author} {\bibfnamefont {P.~A.}\ \bibnamefont {Rikvold}},\
  and\ \bibinfo {author} {\bibfnamefont {Y.-L.}\ \bibnamefont {Wang}},\
  }\bibfield  {title} {\bibinfo {title} {Phase diagram for the
  antiferromagnetic {B}lume-{C}apel model near tricriticality},\ }\href@noop {}
  {\bibfield  {journal} {\bibinfo  {journal} {Phys.\ Rev.\ B}\ }\textbf
  {\bibinfo {volume} {45}},\ \bibinfo {pages} {7237} (\bibinfo {year}
  {1992})}\BibitemShut {NoStop}%
\bibitem [{\citenamefont {Metropolis}\ \emph {et~al.}(1953)\citenamefont
  {Metropolis}, \citenamefont {Rosenbluth}, \citenamefont {Rosenbluth},
  \citenamefont {Teller},\ and\ \citenamefont {Teller}}]{METR53}%
  \BibitemOpen
  \bibfield  {author} {\bibinfo {author} {\bibfnamefont {N.}~\bibnamefont
  {Metropolis}}, \bibinfo {author} {\bibfnamefont {A.~W.}\ \bibnamefont
  {Rosenbluth}}, \bibinfo {author} {\bibfnamefont {M.~N.}\ \bibnamefont
  {Rosenbluth}}, \bibinfo {author} {\bibfnamefont {A.~H.}\ \bibnamefont
  {Teller}},\ and\ \bibinfo {author} {\bibfnamefont {E.}~\bibnamefont
  {Teller}},\ }\bibfield  {title} {\bibinfo {title} {Equation of state
  calculation by fast computing machines},\ }\href@noop {} {\bibfield
  {journal} {\bibinfo  {journal} {J.\ Chem.\ Phys.}\ }\textbf {\bibinfo
  {volume} {21}},\ \bibinfo {pages} {1087} (\bibinfo {year}
  {1953})}\BibitemShut {NoStop}%
\bibitem [{\citenamefont {Janke}(2008)}]{JANK08}%
  \BibitemOpen
  \bibfield  {author} {\bibinfo {author} {\bibfnamefont {W.}~\bibnamefont
  {Janke}},\ }\bibfield  {title} {\bibinfo {title} {Monte {C}arlo metods in
  classical statistical physics},\ }in\ \href@noop {} {\emph {\bibinfo
  {booktitle} {Computational Many-Particle Physics, Lectrue Notes in Physics,
  vol.\ 739}}},\ \bibinfo {editor} {edited by\ \bibinfo {editor} {\bibfnamefont
  {H.}~\bibnamefont {Fehske}}, \bibinfo {editor} {\bibfnamefont
  {R.}~\bibnamefont {Schneider}},\ and\ \bibinfo {editor} {\bibfnamefont
  {A.}~\bibnamefont {Wei{\ss}e}}}\ (\bibinfo  {publisher} {Springer},\ \bibinfo
  {address} {Berlin, Heidelberg},\ \bibinfo {year} {2008})\ pp.\ \bibinfo
  {pages} {79--140}\BibitemShut {NoStop}%
\bibitem [{\citenamefont {Privman}(1990)}]{PRIV90B}%
  \BibitemOpen
  \bibinfo {editor} {\bibfnamefont {V.}~\bibnamefont {Privman}},\ ed.,\
  \href@noop {} {\emph {\bibinfo {title} {Finite-Size Scaling and Numerical
  Simulation of Statistical Systems}}}\ (\bibinfo  {publisher} {World
  Scientific},\ \bibinfo {address} {Singapore},\ \bibinfo {year}
  {1990})\BibitemShut {NoStop}%
\bibitem [{\citenamefont {Binder}(1981)}]{BIND81}%
  \BibitemOpen
  \bibfield  {author} {\bibinfo {author} {\bibfnamefont {K.}~\bibnamefont
  {Binder}},\ }\bibfield  {title} {\bibinfo {title} {Finite size scaling
  analysis of {I}sing model block distribution functions},\ }\href@noop {}
  {\bibfield  {journal} {\bibinfo  {journal} {Z.\ Phys.\ B}\ }\textbf {\bibinfo
  {volume} {43}},\ \bibinfo {pages} {119} (\bibinfo {year} {1981})}\BibitemShut
  {NoStop}%
\bibitem [{\citenamefont {Selke}\ and\ \citenamefont {Shchur}(2005)}]{SELK05}%
  \BibitemOpen
  \bibfield  {author} {\bibinfo {author} {\bibfnamefont {W.}~\bibnamefont
  {Selke}}\ and\ \bibinfo {author} {\bibfnamefont {L.~N.}\ \bibnamefont
  {Shchur}},\ }\bibfield  {title} {\bibinfo {title} {Critical {B}inder cumulant
  in two-dimensional anisotropic {I}sing models},\ }\href@noop {} {\bibfield
  {journal} {\bibinfo  {journal} {J.\ Phys.\ A: Math.\ Gen.}\ }\textbf
  {\bibinfo {volume} {38}},\ \bibinfo {pages} {L739} (\bibinfo {year}
  {2005})}\BibitemShut {NoStop}%
\bibitem [{\citenamefont {Lee}\ and\ \citenamefont
  {Kosterlitz}(1990)}]{JLEE90}%
  \BibitemOpen
  \bibfield  {author} {\bibinfo {author} {\bibfnamefont {J.}~\bibnamefont
  {Lee}}\ and\ \bibinfo {author} {\bibfnamefont {J.~M.}\ \bibnamefont
  {Kosterlitz}},\ }\bibfield  {title} {\bibinfo {title} {New numerical method
  to study phase transitions},\ }\href@noop {} {\bibfield  {journal} {\bibinfo
  {journal} {Phys.\ Rev.\ Lett.}\ }\textbf {\bibinfo {volume} {65}},\ \bibinfo
  {pages} {137} (\bibinfo {year} {1990})}\BibitemShut {NoStop}%
\bibitem [{\citenamefont {Lee}\ and\ \citenamefont
  {Kosterlitz}(1991)}]{JLEE91}%
  \BibitemOpen
  \bibfield  {author} {\bibinfo {author} {\bibfnamefont {J.}~\bibnamefont
  {Lee}}\ and\ \bibinfo {author} {\bibfnamefont {J.~M.}\ \bibnamefont
  {Kosterlitz}},\ }\bibfield  {title} {\bibinfo {title} {Finite-size scaling
  and {M}onte {C}arlo simulations of first-order phase transitions},\
  }\href@noop {} {\bibfield  {journal} {\bibinfo  {journal} {Phys.\ Rev.\ B}\
  }\textbf {\bibinfo {volume} {43}},\ \bibinfo {pages} {3265} (\bibinfo {year}
  {1991})}\BibitemShut {NoStop}%
\bibitem [{\citenamefont {Paw{\l}owski}(2009)}]{PAWL09}%
  \BibitemOpen
  \bibfield  {author} {\bibinfo {author} {\bibfnamefont {G.}~\bibnamefont
  {Paw{\l}owski}},\ }\bibfield  {title} {\bibinfo {title} {Percolation
  properties of the antiferromagnetic {B}lume {C}apel model in the presence of
  a magnetic field},\ }\href@noop {} {\bibfield  {journal} {\bibinfo  {journal}
  {Physica A}\ }\textbf {\bibinfo {volume} {388}},\ \bibinfo {pages} {1111}
  (\bibinfo {year} {2009})}\BibitemShut {NoStop}%
\bibitem [{\citenamefont {Hasenbusch}(2010)}]{HASE10}%
  \BibitemOpen
  \bibfield  {author} {\bibinfo {author} {\bibfnamefont {M.}~\bibnamefont
  {Hasenbusch}},\ }\bibfield  {title} {\bibinfo {title} {Finite size scaling
  study of lattice models in the three-dimensional {I}sing universality
  class},\ }\href@noop {} {\bibfield  {journal} {\bibinfo  {journal} {Phys.\
  Rev.\ B}\ }\textbf {\bibinfo {volume} {82}},\ \bibinfo {pages} {174433}
  (\bibinfo {year} {2010})}\BibitemShut {NoStop}%
\bibitem [{\citenamefont {Ron}\ \emph {et~al.}(2017)\citenamefont {Ron},
  \citenamefont {Brandt},\ and\ \citenamefont {Swendsen}}]{RON17}%
  \BibitemOpen
  \bibfield  {author} {\bibinfo {author} {\bibfnamefont {D.}~\bibnamefont
  {Ron}}, \bibinfo {author} {\bibfnamefont {A.}~\bibnamefont {Brandt}},\ and\
  \bibinfo {author} {\bibfnamefont {R.~H.}\ \bibnamefont {Swendsen}},\
  }\bibfield  {title} {\bibinfo {title} {Surprising convergence of the {M}onte
  {C}arlo renormalization group for the three-dimensional {I}sing model},\
  }\href@noop {} {\bibfield  {journal} {\bibinfo  {journal} {Phys.\ Rev.\ E}\
  }\textbf {\bibinfo {volume} {95}},\ \bibinfo {pages} {053305} (\bibinfo
  {year} {2017})}\BibitemShut {NoStop}%
\bibitem [{\citenamefont {Ferrenberg}\ \emph {et~al.}(2018)\citenamefont
  {Ferrenberg}, \citenamefont {Xu},\ and\ \citenamefont {Landau}}]{FERR18}%
  \BibitemOpen
  \bibfield  {author} {\bibinfo {author} {\bibfnamefont {A.~M.}\ \bibnamefont
  {Ferrenberg}}, \bibinfo {author} {\bibfnamefont {J.}~\bibnamefont {Xu}},\
  and\ \bibinfo {author} {\bibfnamefont {D.~P.}\ \bibnamefont {Landau}},\
  }\bibfield  {title} {\bibinfo {title} {Pushing the limits of {M}onte {C}arlo
  simulations for the three-dimensional {I}sing model},\ }\href@noop {}
  {\bibfield  {journal} {\bibinfo  {journal} {Phys.\ Rev.\ E}\ }\textbf
  {\bibinfo {volume} {97}},\ \bibinfo {pages} {043301} (\bibinfo {year}
  {2018})}\BibitemShut {NoStop}%
\bibitem [{\citenamefont {Xu}\ \emph {et~al.}(2020)\citenamefont {Xu},
  \citenamefont {Ferrenberg},\ and\ \citenamefont {Landau}}]{XU20}%
  \BibitemOpen
  \bibfield  {author} {\bibinfo {author} {\bibfnamefont {J.}~\bibnamefont
  {Xu}}, \bibinfo {author} {\bibfnamefont {A.~M.}\ \bibnamefont {Ferrenberg}},\
  and\ \bibinfo {author} {\bibfnamefont {D.~P.}\ \bibnamefont {Landau}},\
  }\bibfield  {title} {\bibinfo {title} {High-resolution {M}onte {C}arlo study
  of the order-parameter distribution of the three-dimensional {I}sing model},\
  }\href@noop {} {\bibfield  {journal} {\bibinfo  {journal} {Phys.\ Rev.\ E}\
  }\textbf {\bibinfo {volume} {101}},\ \bibinfo {pages} {023315} (\bibinfo
  {year} {2020})}\BibitemShut {NoStop}%
\bibitem [{\citenamefont {Deserno}(1997)}]{DESE97}%
  \BibitemOpen
  \bibfield  {author} {\bibinfo {author} {\bibfnamefont {M.}~\bibnamefont
  {Deserno}},\ }\bibfield  {title} {\bibinfo {title} {Tricriticality and the
  {B}lume-{C}apel model: A {M}onte {C}arlo study within the microcanonical
  ensemble},\ }\href@noop {} {\bibfield  {journal} {\bibinfo  {journal} {Phys.\
  Rev.\ E}\ }\textbf {\bibinfo {volume} {56}},\ \bibinfo {pages} {5204}
  (\bibinfo {year} {1997})}\BibitemShut {NoStop}%
\bibitem [{\citenamefont {Deng}\ and\ \citenamefont
  {Bl{\"o}te}(2004)}]{DENG04}%
  \BibitemOpen
  \bibfield  {author} {\bibinfo {author} {\bibfnamefont {Y.}~\bibnamefont
  {Deng}}\ and\ \bibinfo {author} {\bibfnamefont {H.~W.~J.}\ \bibnamefont
  {Bl{\"o}te}},\ }\bibfield  {title} {\bibinfo {title} {Constrained tricritical
  {B}lume-{C}apel model in three dimensions},\ }\href@noop {} {\bibfield
  {journal} {\bibinfo  {journal} {Phys.\ Rev.\ E}\ }\textbf {\bibinfo {volume}
  {70}},\ \bibinfo {pages} {046111} (\bibinfo {year} {2004})}\BibitemShut
  {NoStop}%
\bibitem [{\citenamefont {Zierenberg}\ \emph {et~al.}(2015)\citenamefont
  {Zierenberg}, \citenamefont {Fytas},\ and\ \citenamefont {Janke}}]{ZIER15}%
  \BibitemOpen
  \bibfield  {author} {\bibinfo {author} {\bibfnamefont {J.}~\bibnamefont
  {Zierenberg}}, \bibinfo {author} {\bibfnamefont {N.~G.}\ \bibnamefont
  {Fytas}},\ and\ \bibinfo {author} {\bibfnamefont {W.}~\bibnamefont {Janke}},\
  }\bibfield  {title} {\bibinfo {title} {Parallel multicanonical study of the
  three-dimensional {B}lune-{C}apel model},\ }\href@noop {} {\bibfield
  {journal} {\bibinfo  {journal} {Phys.\ Rev.\ E}\ }\textbf {\bibinfo {volume}
  {91}},\ \bibinfo {pages} {032126} (\bibinfo {year} {2015})}\BibitemShut
  {NoStop}%
\bibitem [{\citenamefont {Binder}\ and\ \citenamefont {Landau}(1984)}]{BIND84}%
  \BibitemOpen
  \bibfield  {author} {\bibinfo {author} {\bibfnamefont {K.}~\bibnamefont
  {Binder}}\ and\ \bibinfo {author} {\bibfnamefont {D.~P.}\ \bibnamefont
  {Landau}},\ }\bibfield  {title} {\bibinfo {title} {Finite-size scaling at
  first-order phase transitions},\ }\href@noop {} {\bibfield  {journal}
  {\bibinfo  {journal} {Phys.\ Rev.\ B}\ }\textbf {\bibinfo {volume} {30}},\
  \bibinfo {pages} {1477} (\bibinfo {year} {1984})}\BibitemShut {NoStop}%
\bibitem [{\citenamefont {Borgs}\ and\ \citenamefont
  {Koteck{\'y}}(1990)}]{BORG90}%
  \BibitemOpen
  \bibfield  {author} {\bibinfo {author} {\bibfnamefont {C.}~\bibnamefont
  {Borgs}}\ and\ \bibinfo {author} {\bibfnamefont {R.}~\bibnamefont
  {Koteck{\'y}}},\ }\bibfield  {title} {\bibinfo {title} {A rigorous theory of
  finite-size scaling at first-order phase transitions},\ }\href@noop {}
  {\bibfield  {journal} {\bibinfo  {journal} {J.\ Stat.\ Phys.}\ }\textbf
  {\bibinfo {volume} {61}},\ \bibinfo {pages} {79} (\bibinfo {year}
  {1990})}\BibitemShut {NoStop}%
\bibitem [{\citenamefont {Borgs}\ and\ \citenamefont
  {Koteck{\'y}}(1992)}]{BORG92A}%
  \BibitemOpen
  \bibfield  {author} {\bibinfo {author} {\bibfnamefont {C.}~\bibnamefont
  {Borgs}}\ and\ \bibinfo {author} {\bibfnamefont {R.}~\bibnamefont
  {Koteck{\'y}}},\ }\bibfield  {title} {\bibinfo {title} {Finite-size effects
  at asymmetric first-order phase transitions},\ }\href@noop {} {\bibfield
  {journal} {\bibinfo  {journal} {Phys.\ Rev.\ Lett.}\ }\textbf {\bibinfo
  {volume} {68}},\ \bibinfo {pages} {1734} (\bibinfo {year}
  {1992})}\BibitemShut {NoStop}%
\bibitem [{\citenamefont {Borgs}\ and\ \citenamefont
  {Kappler}(1992)}]{BORG92B}%
  \BibitemOpen
  \bibfield  {author} {\bibinfo {author} {\bibfnamefont {C.}~\bibnamefont
  {Borgs}}\ and\ \bibinfo {author} {\bibfnamefont {S.}~\bibnamefont
  {Kappler}},\ }\bibfield  {title} {\bibinfo {title} {Equal weight versus equal
  height: a numerical study of an asymmetric first-order transition},\
  }\href@noop {} {\bibfield  {journal} {\bibinfo  {journal} {Phys.\ Lett.\ A}\
  }\textbf {\bibinfo {volume} {171}},\ \bibinfo {pages} {37} (\bibinfo {year}
  {1992})}\BibitemShut {NoStop}%
\bibitem [{\citenamefont {Vollmayr}\ \emph {et~al.}(1993)\citenamefont
  {Vollmayr}, \citenamefont {Reger}, \citenamefont {Scheucher},\ and\
  \citenamefont {Binder}}]{VOLL93}%
  \BibitemOpen
  \bibfield  {author} {\bibinfo {author} {\bibfnamefont {K.}~\bibnamefont
  {Vollmayr}}, \bibinfo {author} {\bibfnamefont {J.~D.}\ \bibnamefont {Reger}},
  \bibinfo {author} {\bibfnamefont {M.}~\bibnamefont {Scheucher}},\ and\
  \bibinfo {author} {\bibfnamefont {K.}~\bibnamefont {Binder}},\ }\bibfield
  {title} {\bibinfo {title} {Finite-size effects at thermally-driven first
  order phase transitions: a phenomenological theory of the order parameter
  distribution},\ }\href@noop {} {\bibfield  {journal} {\bibinfo  {journal}
  {Z.\ Phys.\ B}\ }\textbf {\bibinfo {volume} {91}},\ \bibinfo {pages} {113}
  (\bibinfo {year} {1993})}\BibitemShut {NoStop}%
\bibitem [{\citenamefont {Tsai}\ and\ \citenamefont {Salinas}(1998)}]{TSAI98}%
  \BibitemOpen
  \bibfield  {author} {\bibinfo {author} {\bibfnamefont {S.-H.}\ \bibnamefont
  {Tsai}}\ and\ \bibinfo {author} {\bibfnamefont {S.~R.}\ \bibnamefont
  {Salinas}},\ }\bibfield  {title} {\bibinfo {title} {Fourth-order cumulants to
  characterize the phase transitions of a spin-1 {I}sing model},\ }\href@noop
  {} {\bibfield  {journal} {\bibinfo  {journal} {Braz.\ J.\ Phys.}\ }\textbf
  {\bibinfo {volume} {28}},\ \bibinfo {pages} {58} (\bibinfo {year}
  {1998})}\BibitemShut {NoStop}%
\bibitem [{\citenamefont {Chan}\ and\ \citenamefont {Rikvold}(2015)}]{CHAN15}%
  \BibitemOpen
  \bibfield  {author} {\bibinfo {author} {\bibfnamefont {C.~H.}\ \bibnamefont
  {Chan}}\ and\ \bibinfo {author} {\bibfnamefont {P.~A.}\ \bibnamefont
  {Rikvold}},\ }\bibfield  {title} {\bibinfo {title} {Monte {C}arlo simulations
  of the critical properties of a {Z}iff-{G}ulari-{B}arshad model of catalytic
  {CO} oxidation with long-range reactivity},\ }\href@noop {} {\bibfield
  {journal} {\bibinfo  {journal} {Phys.\ Rev.\ E}\ }\textbf {\bibinfo {volume}
  {91}},\ \bibinfo {pages} {012103} (\bibinfo {year} {2015})}\BibitemShut
  {NoStop}%
\bibitem [{\citenamefont {Childs}\ \emph {et~al.}(2012)\citenamefont {Childs},
  \citenamefont {Brugger}, \citenamefont {Whitlock}, \citenamefont {Meredith},
  \citenamefont {Ahern}, \citenamefont {Pugmire}, \citenamefont {Biagas},
  \citenamefont {Miller}, \citenamefont {Harrison}, \citenamefont {Weber},
  \citenamefont {Krishnan}, \citenamefont {Fogal}, \citenamefont {Sanderson},
  \citenamefont {Garth}, \citenamefont {Bethel}, \citenamefont {Camp},
  \citenamefont {R\"{u}bel}, \citenamefont {Durant}, \citenamefont {Favre},\
  and\ \citenamefont {Navr\'{a}til}}]{VISIT12}%
  \BibitemOpen
  \bibfield  {author} {\bibinfo {author} {\bibfnamefont {H.}~\bibnamefont
  {Childs}}, \bibinfo {author} {\bibfnamefont {E.}~\bibnamefont {Brugger}},
  \bibinfo {author} {\bibfnamefont {B.}~\bibnamefont {Whitlock}}, \bibinfo
  {author} {\bibfnamefont {J.}~\bibnamefont {Meredith}}, \bibinfo {author}
  {\bibfnamefont {S.}~\bibnamefont {Ahern}}, \bibinfo {author} {\bibfnamefont
  {D.}~\bibnamefont {Pugmire}}, \bibinfo {author} {\bibfnamefont
  {K.}~\bibnamefont {Biagas}}, \bibinfo {author} {\bibfnamefont
  {M.}~\bibnamefont {Miller}}, \bibinfo {author} {\bibfnamefont
  {C.}~\bibnamefont {Harrison}}, \bibinfo {author} {\bibfnamefont {G.~H.}\
  \bibnamefont {Weber}}, \bibinfo {author} {\bibfnamefont {H.}~\bibnamefont
  {Krishnan}}, \bibinfo {author} {\bibfnamefont {T.}~\bibnamefont {Fogal}},
  \bibinfo {author} {\bibfnamefont {A.}~\bibnamefont {Sanderson}}, \bibinfo
  {author} {\bibfnamefont {C.}~\bibnamefont {Garth}}, \bibinfo {author}
  {\bibfnamefont {E.~W.}\ \bibnamefont {Bethel}}, \bibinfo {author}
  {\bibfnamefont {D.}~\bibnamefont {Camp}}, \bibinfo {author} {\bibfnamefont
  {O.}~\bibnamefont {R\"{u}bel}}, \bibinfo {author} {\bibfnamefont
  {M.}~\bibnamefont {Durant}}, \bibinfo {author} {\bibfnamefont {J.~M.}\
  \bibnamefont {Favre}},\ and\ \bibinfo {author} {\bibfnamefont
  {P.}~\bibnamefont {Navr\'{a}til}},\ }\bibfield  {title} {\bibinfo {title}
  {Vis{I}t: An end-user tool for visualizing and analyzing very large data},\
  }in\ \href {https://doi.org/10.1201/b12985} {\emph {\bibinfo {booktitle}
  {High Performance Visualization--Enabling Extreme-Scale Scientific
  Insight}}}\ (\bibinfo {year} {2012})\ pp.\ \bibinfo {pages}
  {357--372}\BibitemShut {NoStop}%
\bibitem [{\citenamefont {Kittel}(2004)}]{KITTEL8}%
  \BibitemOpen
  \bibfield  {author} {\bibinfo {author} {\bibfnamefont {C.}~\bibnamefont
  {Kittel}},\ }\href@noop {} {\emph {\bibinfo {title} {Introduction to Solid
  State Physics, 8th Edition}}}\ (\bibinfo  {publisher} {Wiley},\ \bibinfo
  {address} {Hoboken, NJ},\ \bibinfo {year} {2004})\ \bibinfo {note}
  {{Ch}.~2}\BibitemShut {NoStop}%
\end{thebibliography}

%

\end{document}